%% file: acl2.tex
\documentclass[11pt]{article}

\PassOptionsToPackage{table}{xcolor}

\usepackage[preprint]{acl}

\usepackage{times}
\usepackage{latexsym}

\usepackage[T1]{fontenc}

\usepackage[utf8]{inputenc}

\usepackage{microtype}

\usepackage{inconsolata}

\usepackage{graphicx}

\usepackage{xspace}
\usepackage{amsmath}
\usepackage{amsfonts}
\usepackage{amssymb}
\usepackage{booktabs}
\usepackage{multirow}
\usepackage{makecell}
\usepackage{algorithm}
\usepackage{algpseudocode}
\usepackage{subcaption}
\usepackage{placeins}

\algrenewcommand{\algorithmiccomment}[1]{\hfill$\triangleright$ #1}

\newif\ifdraft
\drafttrue

\ifdraft
  \newcommand{\todocolor}[2]{\textcolor{#1}{#2}}
\else
  \newcommand{\todocolor}[2]{}
\fi

\newcommand{\figref}[1]{Fig.~\ref{#1}}

\newcommand{\tabref}[1]{Tab.~\ref{#1}}
\newcommand{\tabrefs}[2]{Tabs.~\ref{#1}--\ref{#2}}

\newcommand{\appref}[1]{App.~\ref{#1}}

\newcommand{\eqnref}[1]{Eq.~\eqref{#1}}

\title{Black-Box Membership Inference via Word-Level Probability Estimation}

\author{Shengjie Niu\textsuperscript{1}, Yeheng Ge\textsuperscript{1}, \and Jian Huang\textsuperscript{1,2}\thanks{Corresponding author.} \\
\textsuperscript{1}Department of Data Science and Artificial Intelligence \\
The Hong Kong Polytechnic University, Hong Kong SAR, China \\
\textsuperscript{2}Department of Applied Mathematics \\
The Hong Kong Polytechnic University, Hong Kong SAR, China \\
        \texttt{shengjie.niu@connect.polyu.hk, \{yeheng.ge,j.huang\}@polyu.edu.hk}
        }

\begin{document}
\maketitle
\begin{abstract}
Membership inference attacks (MIAs) have emerged as critical tools for auditing privacy risks in large language models (LLMs), aiming to determine whether a given text was included in a model's training corpus. However, most existing MIAs require access to per-token logits or probabilities, making them inapplicable in practice to proprietary LLMs that expose only textual continuations. To address this underexplored setting, we propose Word-level Probability MIA (WPMIA), a statistically principled MIA for strict black-box privacy auditing. WPMIA estimates word-level generation probabilities via Monte Carlo sampling with local kernel smoothing, then aggregates these estimates into a sequence-level likelihood estimator. Furthermore, WPMIA constructs the likelihood conditioned on different prefixes, thereby amplifying the distributional differences between members and non-members. We evaluate WPMIA across various open-source LLMs and find that it consistently outperforms existing black-box baselines. Importantly, we also evaluate WPMIA on modern proprietary LLMs, including GPT-5-Chat, Gemini-2.5-Flash, and Claude-4.5-Haiku, achieving an average TPR@5\%FPR of 42.0 across these models.
These results offer a sound foundation for future research on strict black-box membership inference. Code is available at \href{https://github.com/niusj03/WPMIA}{https://github.com/niusj03/WPMIA}.
\end{abstract}

\input{sec/1_intro}
\input{sec/2_method}
\input{sec/3_exp}
\input{sec/4_conclusion}

\section*{Acknowledgments}
Jian Huang acknowledges support from The Hong Kong Polytechnic University (Research Grant P0046811).
We used ChatGPT solely for language polishing and refinement of the manuscript. It was not used to generate research ideas, design experiments, analyze results, or formulate conclusions. All final text was reviewed and verified by the authors.

\bibliography{acl2}

\input{sec/X_appendix}

\end{document}

%% file: sec/1_intro.tex
\section{Introduction}\label{sec:intro}

\begin{figure}[t!]
    \centering
    \includegraphics[width=1\linewidth]{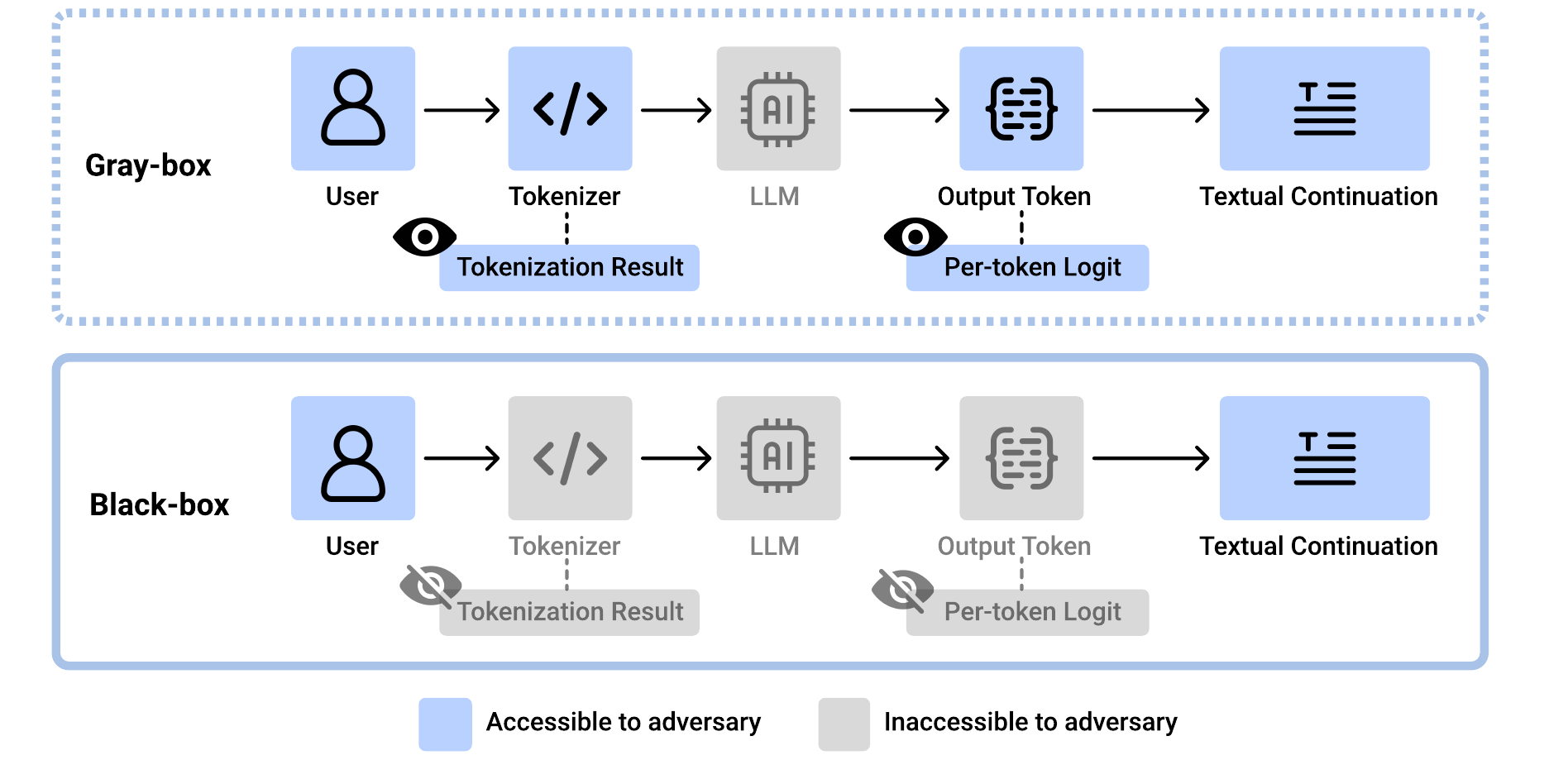}
    \vspace{-1.5em}
    \caption{\textbf{Adversary access levels.} \emph{Gray-box}: the adversary has access to tokenization results and per-token logits, but not to the model's internal weights. \emph{Black-box (of interest)}: the adversary has no access to model weights, tokenization results, or token logits, and can only observe textual continuations.}
    \label{fig:access_level}
    \vspace{-1em}
\end{figure}

As the pretraining corpora of modern large language models (LLMs) continue to grow in scale~\citep{raffel2020exploring,brown2020language}, they may still contain sensitive or restricted content despite data filtering~\citep{almazrouei2023falcon}.
This raises concerns about the inclusion of copyrighted content~\citep{chang2023speak,duarte2024decop} and personally identifiable information~\citep{mozes2023use,tang2024privacy-preserving}.
To audit these risks, pretraining data detection~\citep{mink} aims to determine whether a given text was included in the training corpus of a target LLM, typically through membership inference attacks (MIAs)~\citep{mia}.
Prior studies have developed effective MIA methods for LLMs~\citep{reference,neighborhood,mink,mink+}.
However, most of these methods require \emph{gray-box} access, where the adversary can obtain tokenization results and per-token logits, as shown in \figref{fig:access_level}.
In practical deployment, proprietary LLMs such as ChatGPT and Claude are typically accessed through APIs or subscriptions, where users can only observe textual continuations.
Consequently, these gray-box methods are inapplicable to \emph{black-box} proprietary LLMs that do not disclose per-token logits.
Despite its practical relevance, black-box pretraining data detection remains underexplored.

MIA methods generally rely on the overfitting effect that a text included in the training data (\emph{member}) tends to have a higher likelihood than a text not seen during training (\emph{non-member})~\cite{loss}.
Given a target text split into a sequence of tokens, gray-box MIA methods can directly compute sequence-level likelihoods from token-level logits~\citep{revisit_mia}. 
Based on this likelihood signal, they further improve detection by focusing on low-probability tokens~\citep{mink,mink+}, measuring likelihood changes under textual perturbations~\citep{neighborhood,recall,conrecall}.
In contrast, recent black-box methods~\citep{petal,samia,simmia} cannot directly compute likelihoods and instead attempt to recover membership signals from generated continuations alone.
However, these methods are developed largely through heuristics rather than a principled statistical formulation of the membership signal.
Therefore, constructing a reliable and theoretically grounded black-box MIA remains an open problem.

We propose Word-level Probability MIA (WPMIA), a statistically principled MIA for the black-box setting.
The key insight is to recover the likelihood signal directly from textual continuations.
Given a target text represented as a sequence of words, WPMIA estimates the generation probability of each word from Monte Carlo samples~\citep{mc} of the target model's continuations, and aggregates these word-level probabilities into a sequence-level likelihood estimator.
Moreover, WPMIA constructs both member and non-member prefixes and estimates the corresponding prefix-conditioned sequence-level likelihoods.
It then contrasts the likelihoods conditioned on different prefixes, thereby amplifying the membership signal.
Compared with existing black-box MIAs, WPMIA directly estimates the sequence-level likelihood signal, providing a statistically sound alternative to heuristic black-box scoring.
Importantly, when applying prefix-based enhancement, WPMIA first aggregates word-level probabilities into sequence-level likelihoods and then contrasts the conditional likelihood, avoiding the unstable word-level ratios.

We conduct extensive experiments on three standard benchmarks, WikiMIA~\citep{mink}, MIMIR~\citep{mimir}, and WikiMIA-25~\citep{simmia}, covering a broad range of open-source LLM families.
Importantly, we also evaluate recent proprietary LLMs, including GPT-5-Chat~\citep{gpt5}, Gemini-2.5-Flash~\citep{gemini25}, and Claude-4.5-Haiku~\citep{claudeharku}.
\textbf{(1)} On open-source LLMs, WPMIA achieves the strongest black-box MIA performance and matches or exceeds many gray-box MIAs.
\textbf{(2)} On proprietary LLMs, WPMIA remains effective, achieving an average TPR@5\%FPR of 42.0 across commercial models, which validates the practical feasibility of black-box privacy auditing for commercial LLMs without access to the target tokenizer or per-token logits.

%% file: sec/2_method.tex
\section{Method}\label{sec:method}

\subsection{Preliminary}\label{sec:method_prelimi}

\paragraph{Membership Inference against LLMs.}
Let $f_{\theta}$ denote an autoregressive LLM parameterized by $\theta$ and trained on a corpus $D$. 
Denote a text as a sequence of tokens $x=(x_1,\ldots,x_L)$, where $x_{<i}:=(x_1,\ldots,x_{i-1})$ denotes the preceding tokens before $x_i$.
Given $x_{<i}$, the model defines a conditional distribution $f_{\theta}(\cdot \mid x_{<i})$ over the vocabulary, and $f_{\theta}(x_i \mid x_{<i})$ denotes the probability assigned to the next token $x_i$.
Membership inference attacks (MIAs) aim to determine whether a text $x$ was included in $D$.
By the chain rule, the LLM assigns $x$ the following probability:
\begin{equation}
p_{\theta}(x) = \prod_{i=1}^{L} f_{\theta}(x_i \mid x_{<i}).
\end{equation}
In practice, the average log-likelihood (LL) is used for numerical stability:
\begin{equation}
\label{eq:avg_ll}
LL(x) = \frac{1}{L}\sum_{t=1}^{L} \log f_{\theta}(x_t \mid x_{<t}).
\end{equation}
LL reflects the probability that the target model generates $x$ token-by-token, and is used as a membership signal by many MIA methods.
Generally, an MIA method assigns each text $x$ a membership score $\mathcal{S}(\theta,x)$ based on the log-likelihood $LL(x)$ and decides the membership bit by thresholding this score with a decision threshold $\kappa$:
\begin{equation}
    \mathcal{A}(\theta,x) = \mathbb{I}[\mathcal{S}(\theta,x) > \kappa],
\end{equation}
where $\mathbb{I}[\cdot]$ is the indicator function.
Despite their effectiveness, these likelihood-based MIAs require access to the token-level logits or probabilities when querying the target model $f_\theta$, which are unavailable for many proprietary LLMs.

\paragraph{Threat Model.}
The threat model specifies the information available to the adversary when querying the target model.
Most existing MIAs assume a \emph{gray-box} setting~\citep{reference, mink, mink+, recall, conrecall}, where the adversary has access to tokenization results and per-token logits, but not to the model's internal weights.
We consider a more realistic \emph{black-box} setting, where the adversary has no access to model weights, tokenization results, or logits, and can only observe generated textual continuations~\citep{samia}.
This setting aligns with API interactions for proprietary LLMs, such as ChatGPT, Gemini, and Claude.

\subsection{Estimating Sequence-level Likelihood}\label{sec:estimation}
The ground-truth likelihood of the target text is unavailable in the black-box setting, so we first aim to estimate its sequence-level likelihood.
Moreover, since the target model's tokenizer is unavailable, we consider words rather than model-specific tokens as the basic units, and split the text into a word sequence $x=(x_1,\ldots,x_L)$ using a standard word tokenizer~\citep{bird2009natural}.
We start from word-level primitives and work on the log scale, so that the resulting score preserves the additive structure of log-likelihoods in \eqnref{eq:avg_ll}.
For each word $x_i$, the word-level log-probability is
\begin{equation}
\label{eq:word_ll}
    \widehat{\ell}(x_i | c_i) = \log \left( \epsilon + \widehat{p}(x_i|c_i) \right),
\end{equation}
where $\widehat{p}(x_i | c_i)$ denotes the estimated probability of generating $x_i$ conditioned on the context $c_i$, and $\epsilon>0$ prevents zero-valued estimates.

We use Monte Carlo sampling~\cite{mc} to estimate $\widehat{p}(x_i | c_i)$.
Specifically, for a context $c_i$, we query $f_{\theta}$ $M$ times and collect the \emph{first} generated word from each response.
Let $\widehat{x}_{i,j}^{c_i}$ denote the first word generated in the $j$-th query conditioned on $c_i$.
We estimate the word-level probability mass assigned to $x_i$ by
\begin{equation}
\label{eq:word_prob}
    \widehat{p}(x_i|c_i) = \frac{1}{M} \sum_{j=1}^{M} K(x_i, \widehat{x}_{i,j}^{c_i}),
\end{equation}
where $M$ is the number of Monte Carlo samples and $K$ is a kernel function that assigns probability mass to $x_i$ according to the sampled word $\widehat{x}_{i,j}^{c_i}$.

A straightforward choice is the exact-match kernel: $K_{\mathrm{em}}(x_i,\widehat{x}) = \mathbb{I}[x_i = \widehat{x}]$, under which \eqnref{eq:word_prob} becomes the empirical probability of sampling $x_i$ as the next word.
However, with a limited sampling budget and a large vocabulary, exact matching can lead to sparse or near-zero estimates.
We therefore use a semantic kernel
\begin{equation}
\label{eq:semantic_kernel}
    K_{\tau}(x_i, \widehat{x}) = \exp\left(\frac{\cos(e(x_i), e(\widehat{x}))-1}{\tau}\right),
\end{equation}
where $e(\cdot)$ denotes a word embedding function and $\tau$ is a temperature parameter that controls the smoothness of the semantic kernel.
This kernel estimates the probability of $x_i$ by locally averaging over generated words in the embedding space: words that are semantically closer to $x_i$ receive larger weights, while distant words contribute less.
This local smoothing reduces the sparsity of exact matching and produces more stable probability estimates under a limited sampling budget~\citep{intro_nonpara}.

This formulation has two key advantages.
First, the semantic kernel provides semantic robustness: not only exact matches but also semantically related words can provide evidence for the target word.
Second, the log-scale estimator in \eqnref{eq:word_ll} enables additive aggregation over words, yielding the sequence-level log-likelihood estimator
\begin{equation}
\label{eq:estimated_seq_ll}
    \widehat{LL}^{0}(x) = \frac{1}{L}\sum_{i=1}^{L} \widehat{\ell}(x_i|c_i^{0}),
    \quad c_i^{0}=x_{<i},
\end{equation}
where $\widehat{\ell}(x_i|c_i^{0})$ is the word-level log-probability in \eqnref{eq:word_ll}.
This form aligns with the token-sequence likelihood formulation in \eqnref{eq:avg_ll}.

\subsection{Enhancing with Contrastive Prefixes}\label{sec:method_prefix}

\begin{figure*}[t]
\centering
\includegraphics[width=1\textwidth]{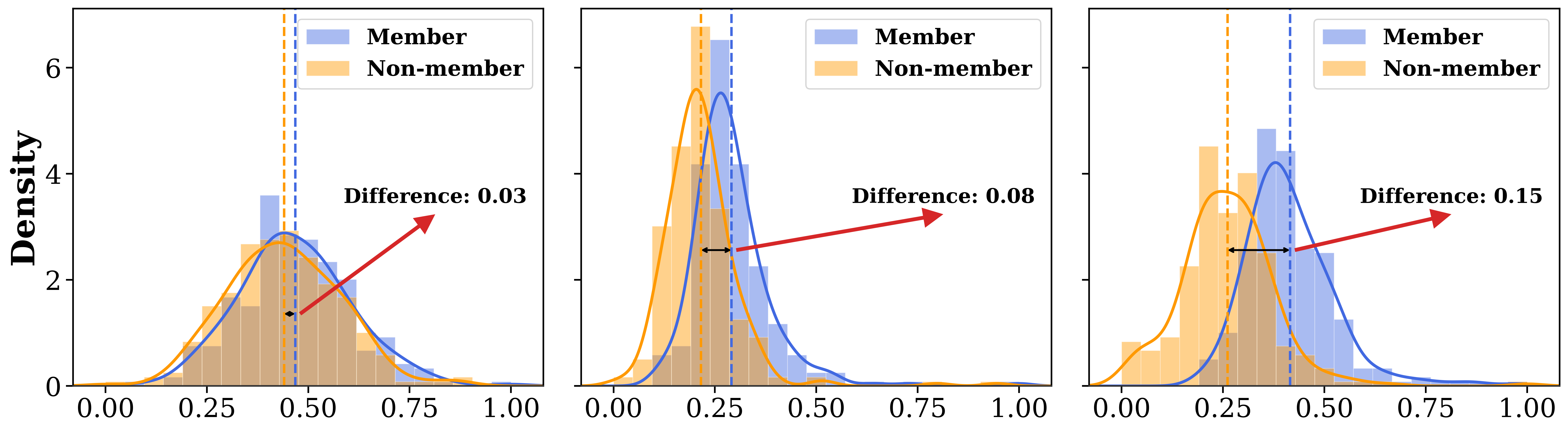}
\caption{\textbf{Visualization of normalized membership-score distributions.}
From left to right, we show min--max-normalized distributions of the unprefixed score $\mathcal{S}_{0}=\widehat{LL}^{0}(x)$, the non-member-prefix relative score $\mathcal{S}_{nm}=\widehat{LL}^{nm}(x)/\widehat{LL}^{0}(x)$, and the contrastive-prefix score $\mathcal{S}_{\mathrm{ctr}}=(\widehat{LL}^{nm}(x)-\widehat{LL}^{m}(x))/\widehat{LL}^{0}(x)$.
The contrastive-prefix score yields the largest separation between members and non-members.}
\label{fig:distribution}
\end{figure*}

Motivated by ReCaLL~\citep{recall} and Con-ReCall~\citep{conrecall}, we view prefix conditioning as a test-time intervention that induces distinct log-likelihood shifts for member and non-member texts.
We use these shifts to strengthen the membership signal through relative comparison.
Specifically, we concatenate either a non-member prefix or a member prefix with the preceding words of the target text.
For $q\in\{nm,m\}$, where $nm$ and $m$ denote non-member and member prefix types, respectively, we define
\begin{equation}
\label{eq:prefix_context}
    c_i^{q} = P_q \oplus x_{<i},
\end{equation}
where $P_{nm}$ and $P_m$ denote non-member and member prefixes, respectively.
Each prefix is constructed by concatenating $T$ texts:
$P_q = p_1^{q} \oplus p_2^{q} \oplus \cdots \oplus p_T^{q}$ for $q\in\{nm,m\}$.

Given the prefix-conditioned context $c_i^q$, we define the prefix-conditioned sequence-level LL estimator as $\widehat{LL}^{q}(x) = \frac{1}{L} \sum_{i=1}^{L}\widehat{\ell}(x_i|c_i^{q})$.
We then use the normalized contrastive LL as the membership score:
\begin{equation}
\label{eq:contrastive_score}
    \mathcal{S}(\theta,x) = \frac{\widehat{LL}^{nm}(x) - \widehat{LL}^{m}(x)}{\widehat{LL}^{0}(x)},
\end{equation}
where $\widehat{LL}^{0}(x)$ is the unprefixed sequence-level log-likelihood estimator in \eqnref{eq:estimated_seq_ll}.
A larger value of $\mathcal{S}(\theta,x)$ indicates stronger evidence of membership.

\figref{fig:distribution} illustrates how the membership score distribution changes as prefix conditioning and contrastive comparison are introduced.
Compared with the unprefixed score $\mathcal{S}_{0}$, the non-member-prefix relative score $\mathcal{S}_{nm}$ increases the separation between members and non-members, while the contrastive-prefix score $\mathcal{S}_{\mathrm{ctr}}$ achieves the largest separation.
This pattern is consistent with the prefix-induced likelihood shifts observed in prior gray-box settings~\citep{conrecall}.
When conditioned on non-member prefixes, member texts tend to experience a larger likelihood reduction than non-member texts.
In contrast, member-prefix conditioning induces asymmetric likelihood shifts between member and non-member texts.
Therefore, \eqnref{eq:contrastive_score} contrasts the likelihoods induced by non-member and member prefixes to enlarge the membership signal, and normalizes this difference by the unprefixed likelihood to account for the base likelihood of the target text.

\paragraph{Remark.}
Recent work~\citep{simmia} computes prefix-conditional scores at the word level and then aggregates these word-level scores into a sequence-level score. In our notation, this strategy takes the form
\begin{equation}
    \label{eq:contrastive_score_word}
    \frac{1}{L}\sum_{i=1}^{L}
    \frac{\widehat{\ell}(x_i|c_i^{nm}) - \widehat{\ell}(x_i|c_i^{m})}
    {\widehat{\ell}(x_i|c_i^{0})}.
\end{equation}
However, we argue that the word-level log-probabilities should first be aggregated into sequence-level LL, and the relative conditional LL should then be computed at the sequence level.
This is because individual word-level log-probabilities can be close to zero, causing unstable ratios, and such per-word normalization breaks the additive structure of sequence-level log-likelihoods. 
We empirically validate this design choice in \tabref{tab:wikimia-pythia-component-wise}.
Algorithm~\ref{alg:wpmia} presents the full pipeline of the proposed method.

\section{Comparison with Heuristic Methods} \label{sec:comparison}

In this section, we compare our likelihood-oriented formulation with existing black-box MIAs that rely on heuristic scoring rather than principled log-likelihood estimation.

\paragraph{SaMIA.}
SaMIA~\citep{samia} approximates a continuation-level pseudo-probability mass.
It divides each target text $x$ into a prefix $x_{\mathrm{prefix}}$ and a reference suffix $x_{\mathrm{suffix}}$.
The LLM then generates $M$ candidate continuations $\{x_{\mathrm{cand}}^{(j)}\}_{j=1}^{M}$ conditioned on $x_{\mathrm{prefix}}$, and SaMIA computes the average ROUGE-N overlap~\citep{rouge} between the continuations and the reference suffix:
\begin{equation*}
\label{eq:samia}
    S_{\mathrm{SaMIA}}(x) = \frac{1}{M} \sum_{j=1}^{M}
    \mathrm{ROUGE}_N(x_{\mathrm{cand}}^{(j)}, x_{\mathrm{suffix}}).
\end{equation*}
This score estimates the expected surface overlap between sampled continuations and the reference suffix.
It can be viewed as a continuation-level pseudo-probability mass, but it is not a log-likelihood estimator and does not decompose into additive word-level log-probabilities.
Moreover, the continuation-level score can suffer from \emph{distribution shift}~\citep{simmia}: as generation proceeds, later words depend on model-generated history rather than the original target context.

\paragraph{SimMIA.}
SimMIA~\citep{simmia} constructs word-level scores and aggregates them into an overall membership score:
\begin{equation}
\label{eq:simmia}
    S_{\mathrm{SimMIA}}(x) = -\frac{1}{L} \sum_{i=1}^{L} \frac{s(x_i \mid c_i^{nm})}{s(x_i \mid c_i^0)}.
\end{equation}
SimMIA considers both hard and soft word-level scores.
The hard score is defined as
\begin{equation}
\label{eq:simmia_hard}
    s^{*}(x_i \mid c_{i}) = \frac{1}{M} \sum_{j=1}^{M}
    \mathbb{I}[x_i = \widehat{x}_{i}^{(j)}],
\end{equation}
which approximates the empirical word-level probability, but is not on a log scale.
The soft score incorporates semantic similarity:
\begin{equation}
\label{eq:simmia_soft}
    s(x_i \mid c_{i}) = \frac{1}{M} \sum_{j=1}^{M} \mathrm{sim}(x_i, \widehat{x}_{i}^{(j)}),
\end{equation}
where $\text{sim}(\cdot,\cdot)$ is the cosine similarity between word embeddings.
Because $s$ is either an empirical frequency or an embedding-based similarity score, the aggregation in \eqnref{eq:simmia} operates on raw score ratios rather than log-probabilities.
Consequently, the resulting average is a heuristic relative score, rather than a sequence-level log-likelihood estimator.

In contrast, our method estimates word-level probability mass and aggregates these word-level estimates into a sequence-level log-likelihood before applying contrastive prefix normalization.
This preserves the additive property of log-likelihoods and avoids the numerical instability of directly averaging word-level scores, where the denominator in \eqnref{eq:simmia} may approach zero.

%% file: sec/3_exp.tex
\section{Experiments}\label{sec:exp}

\subsection{Setup}\label{sec:exp_setup}

\input{tabs/wikimia_bb}

\paragraph{Benchmark.}
We evaluate WPMIA on three widely used benchmarks: WikiMIA~\citep{mink}, MIMIR~\citep{mimir}, and WikiMIA-25~\citep{simmia}.
For WikiMIA, we consider the 32-, 64-, and 128-token sequence-length subsets.
For MIMIR, we use the 7-gram setting and report results across all seven domains: Wikipedia, GitHub, Pile-CC, PubMed Central, ArXiv, DM Mathematics, and HackerNews.
For WikiMIA-25, we follow the cost-efficient evaluation setting of \citet{simmia} by using the length-32 split and sampling a balanced subset of 120 members and 120 non-members.
See detailed dataset statistics in \appref{app:dataset}.

\paragraph{Evaluation Metrics.}
We evaluate MIA methods using the true positive rate (TPR) and false positive rate (FPR). 
We sweep the decision threshold to construct the receiver operating characteristic (ROC) curve, which captures the trade-off between TPR and FPR, and report the area under the ROC curve (AUC) as the primary metric. 
We also report TPR at 5\% FPR to measure attack performance in the low-false-positive regime. 
Unless otherwise specified, averaged results are reported as macro-averages over the corresponding benchmark splits and target models.

\paragraph{Models.}
For WikiMIA, we evaluate four open-source models: OPT-6.7B~\cite{opt}, Pythia-6.9B~\citep{pythia}, LLaMA-13B~\cite{llama}, and GPT-NeoX-20B~\citep{gptneox}. 
For MIMIR, we evaluate the open-source Pythia-dedup family, with model sizes of 160M, 1.4B, 2.8B, and 6.9B. 
For WikiMIA-25, we evaluate the advanced open-source model Qwen3-8B-Base~\cite{qwen3}, along with proprietary LLMs such as Claude-4.5-Haiku, Gemini-2.5-Flash, and GPT-5-Chat.
All open-source models are accessed via Hugging Face, while proprietary models are accessed via their official APIs. 
We follow \citet{simmia} and use the prompt template for next-word sampling on proprietary models.

\paragraph{MIA Baselines.}
We compare WPMIA against representative MIA baselines under both gray-box and black-box access settings.
The gray-box baselines consist of Loss, Reference, Lowercase, Zlib, Neighbor, Min-K\%, Min-K\%++, and ReCaLL.
Since these methods rely on access to the target model's tokenizer and per-token logits, we include their results in the Appendix for reference.
For black-box baselines, we compare against PETAL~\cite{petal}, SaMIA~\citep{samia}, and SimMIA~\citep{simmia}.

\input{tabs/wikimia-25-bb}

\paragraph{Implementation Details.}
We reproduce all baselines using their official hyperparameter configurations.
For WPMIA, we split each target text into a word sequence using the standard word tokenizer~\citep{bird2009natural}.
We set the number of Monte Carlo samples to $M=100$ for WikiMIA and MIMIR, and reduce it to $M=10$ for WikiMIA-25 to control the API cost.
The number of prefix shots is set to $T=7$ for WikiMIA and WikiMIA-25, and $T=10$ for MIMIR.
For each benchmark, prefix examples are held out from the evaluation set for all methods.
We use all-MiniLM-L6-v2~\cite{allminilm} as the embedding model for the semantic kernel.
For the DM Mathematics split in MIMIR, we use exact matching for numerical tokens rather than semantic similarity, since numerical proximity does not necessarily imply semantic equivalence.
Detailed data handling and hyperparameter settings are reported in \appref{app:implementation}.
All open-source experiments are conducted on 8 NVIDIA L40 GPUs, whereas proprietary models are evaluated via their official APIs.

\subsection{Main Results}\label{sec:exp_result}

Due to space constraints, we report black-box comparisons on WikiMIA and WikiMIA-25 in the main manuscript and defer complete results to \appref{app:results}.

\textbf{WPMIA achieves the strongest black-box performance and is competitive with gray-box methods on open-source LLMs.}
As shown in Tabs.~\ref{tab:wikimia_bb} and~\ref{tab:mimir}, WPMIA consistently outperforms existing black-box baselines on both WikiMIA and MIMIR.
On WikiMIA, WPMIA improves over the best black-box baseline by 4.6 AUC points and 14.1 TPR@5\%FPR points on average across the three lengths.
Notably, although WPMIA observes only textual continuations, it also surpasses the strongest gray-box method in several settings, including OPT-6.7B across all WikiMIA lengths, as shown in the complete results in \tabref{tab:wikimia}.
On MIMIR, WPMIA remains the best-performing black-box method across most domain-model-size pairs, showing that its effectiveness generalizes beyond WikiMIA-style data.

\textbf{WPMIA remains effective on advanced open-source and proprietary LLMs.}
As shown in \tabref{tab:wikimia25_bb}, WPMIA achieves strong results on both Pythia-6.9B and the more recent Qwen3-8B-Base, reaching 89.0 and 65.9 AUC, respectively.
More importantly, WPMIA also succeeds on proprietary LLMs where gray-box attacks are inapplicable.
For example, WPMIA achieves AUCs of 66.3, 78.2, and 91.0 on Claude-4.5-Haiku, Gemini-2.5-Flash, and GPT-5-Chat, respectively, with corresponding TPR@5\%FPR scores of 22.1, 36.5, and 67.3.
The complete results, including gray-box baselines for open-source models, are reported in \tabref{tab:wikimia25}.
These results demonstrate the practical feasibility of auditing commercial LLMs under strict black-box access, without relying on the target tokenizer or per-token logits.

\subsection{Analysis}\label{sec:exp_analysis}

We present a detailed analysis of WPMIA in this section.
Unless otherwise specified, we report AUC results on WikiMIA with Pythia-6.9B.

\input{tabs/abla_overall}

\paragraph{Component-Wise Analysis.}
We analyze the contribution of each component of WPMIA and report the results in \tabref{tab:wikimia-pythia-component-wise}.
The baseline uses the unprefixed sequence-level LL estimator as a membership score $\mathcal{S}_{0}=\widehat{LL}^{0}(x)$.
The +NP rows introduce the non-member prefix and use the relative score $\mathcal{S}_{nm}=\widehat{LL}^{nm}(x)/\widehat{LL}^{0}(x)$.
The +MP rows further introduce the member prefix and use the contrastive score $\mathcal{S}_{\mathrm{ctr}}=(\widehat{LL}^{nm}(x)-\widehat{LL}^{m}(x))/\widehat{LL}^{0}(x)$.
For both +NP and +MP, we compare sequence-level normalization with the corresponding word-level normalization described in \eqnref{eq:contrastive_score_word}.

As shown in \tabref{tab:wikimia-pythia-component-wise}, we have two observations.
(1) \textbf{Prefix conditioning amplifies the membership signal beyond the unprefixed likelihood.}
The unprefixed score $\mathcal{S}_{0}$ yields limited AUC scores, e.g., 55.4 at Len. 64.
Introducing the non-member prefix improves the AUC from 55.4 to 77.2, and adding the member prefix further improves it to 86.5, showing that non-member and member prefixes provide complementary likelihood shifts that strengthen the membership signal.
(2) \textbf{Relative conditional likelihood is more effective when computed at the sequence level.}
Although word-level probabilities are useful primitives for likelihood estimation, applying relative normalization before sequence-level likelihood aggregation severely hurts performance, with AUC scores lower than those of the unprefixed LL across all lengths.

\input{tabs/abla_component_analysis}

\paragraph{Crossed Ablation of Estimator and Prefix Conditioning.}
To separate the contributions of word-level estimation and prefix conditioning, \tabref{tab:wikimia_pythia_estimator_ablation} evaluates all combinations of three estimators, namely exact matching (\eqnref{eq:simmia_hard}), raw embedding similarity (\eqnref{eq:simmia_soft}), and the semantic kernel (\eqnref{eq:semantic_kernel}), with three prefix settings.
Two consistent patterns emerge. First, prefix conditioning improves all three estimators: averaged over estimators and sequence lengths, non-member-prefix conditioning raises AUC from 59.1 to 73.3, while introducing member-prefix contrast further improves it to 81.8.
Second, the estimator remains critical under the full contrastive setting. The semantic kernel achieves an average AUC of 85.3, outperforming raw embedding similarity and exact matching by 4.1 and 6.3 points, respectively. 
These results show that contrastive prefix conditioning and semantic-kernel estimation provide complementary gains.

\input{tabs/abla_estimator}

\paragraph{Impact of Continuation Ratio.}
We vary the continuation ratio, the proportion of target text used for membership inference, and examine its effect on attack performance.
As shown in \figref{fig:prefix-ratio}, the AUC of all methods increases as the continuation ratio grows from 0.1 to 0.9.
This trend is expected, since a larger continuation ratio provides a longer target segment and thus more exploitable membership evidence.
WPMIA consistently outperforms SimMIA$^*$ and SimMIA across all continuation ratios, demonstrating its stronger ability to extract membership signals from black-box continuations.

\paragraph{Impact of Sample Size.}
We vary the number of Monte Carlo samples $M$ to study the trade-off between the sampling budget and attack performance.
As shown in \figref{fig:num-shots}, all methods benefit from increasing the number of samples, since more samples reduce the estimation variance of the word-level probability in \eqnref{eq:word_prob}.
WPMIA consistently outperforms both SimMIA$^*$ and SimMIA across all sampling budgets.
Moreover, WPMIA shows better scalability with respect to the sampling budget.
SimMIA saturates early, around $ M=50$, whereas WPMIA continues to improve as $M$ increases and stabilizes after $M=70$, reaching an AUC of around 85.
SimMIA$^*$ improves with more samples but remains consistently below WPMIA.

\paragraph{Impact of Prefix Shots.}
We vary the number of shots $T$ used to construct the member and non-member prefixes.
As shown in \figref{fig:num-samples}, the performance is unstable when only a few shots are used, since 1- or 2-shot prefixes cannot reliably characterize the prefix-conditioned likelihood shifts.
From $T=3$ onward, increasing the number of shots generally improves attack performance, and the results stabilize around $T=8$.
In this range, WPMIA achieves the highest AUC and consistently outperforms SimMIA$^*$ and SimMIA.
These results indicate that a moderate number of prefix shots is sufficient for reliable estimation, and that WPMIA benefits more from additional prefix examples than the baselines.

\input{tabs/abla_embedding}

\paragraph{Impact of Embedding Model.}
To study WPMIA's sensitivity to the choice of embedding model, we evaluate it with two static embedding models, Word2Vec~\citep{Word2Vec} and fastText~\citep{fasttext}, and four dense embedding models: bge-large-en-v1.5~\citep{bge}, all-MiniLM-L6-v2, UAE-Large-V1~\citep{uae}, and mxbai-embed-large-v1~\citep{mxbai}.
As shown in \tabref{tab:wikimia_embedding_ablation}, Word2Vec yields substantially lower AUC, whereas fastText provides a clear improvement across all lengths.
Modern dense embedding models perform strongly and consistently, with average AUCs ranging only from 85.3 to 86.6. This suggests that high-quality semantic matching is important for WPMIA, but its performance is not overly sensitive to the choice of a specific dense embedding model.

In terms of encoder scale, Word2Vec and fastText use 300-dimensional word embeddings. Among the dense encoders, all-MiniLM-L6-v2 produces 384-dimensional embeddings with 22.7M parameters, whereas the other three produce 1,024-dimensional embeddings with approximately 335M parameters each. 
Despite using roughly one-fifteenth as many parameters, all-MiniLM-L6-v2 achieves an average AUC of 85.3, only 1.3 points below the best-performing dense encoder. This result suggests that the default encoder offers a favorable trade-off between attack performance and model scale.

\begin{figure}[h]
    \centering
    \includegraphics[width=0.8\linewidth]{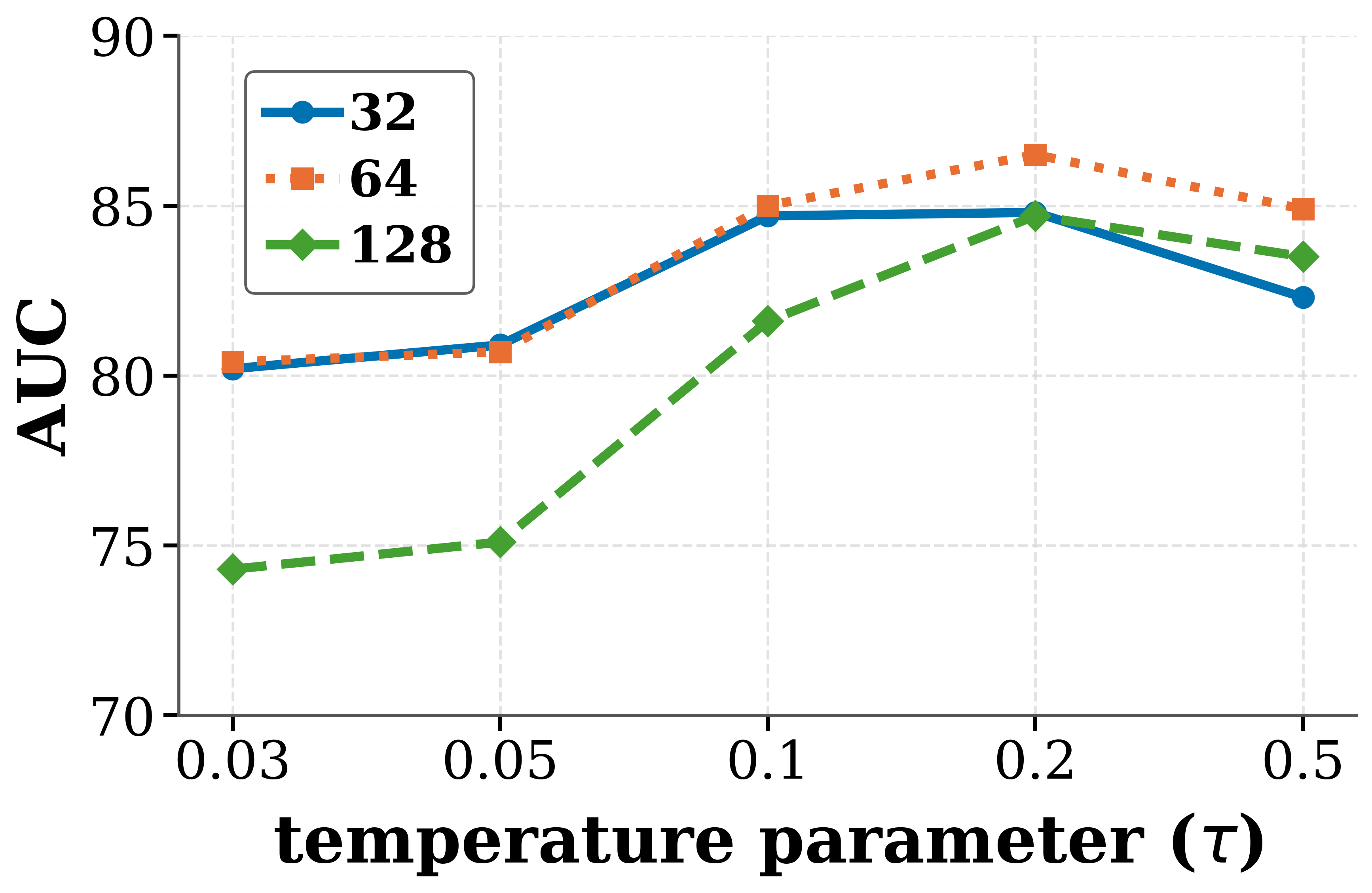}
    \caption{Attack performance under different semantic-kernel temperature parameters $\tau$.}
    \label{fig:abla_tau}
    \vspace{-10pt}
\end{figure}

\input{tabs/abla_robustness_prefix}

\paragraph{Sensitivity to Temperature.}
We analyze the sensitivity of WPMIA to the temperature parameter $\tau$, which controls the smoothness of the semantic kernel in \eqnref{eq:semantic_kernel}.
As shown in \figref{fig:abla_tau}, very small temperatures lead to weaker performance, especially at Len. 128.
This is because an overly sharp kernel may discard useful semantic evidence from generated words that are not exact matches but remain semantically close to the target word.
Increasing $\tau$ improves AUC across all lengths, with the near-best performance obtained around $\tau=0.2$.
When $\tau$ further increases to $0.5$, the performance slightly drops, suggesting that excessive smoothing weakens the distinction between informative and uninformative matches.

\begin{figure}[h]
    \centering
    \includegraphics[width=0.8\linewidth]{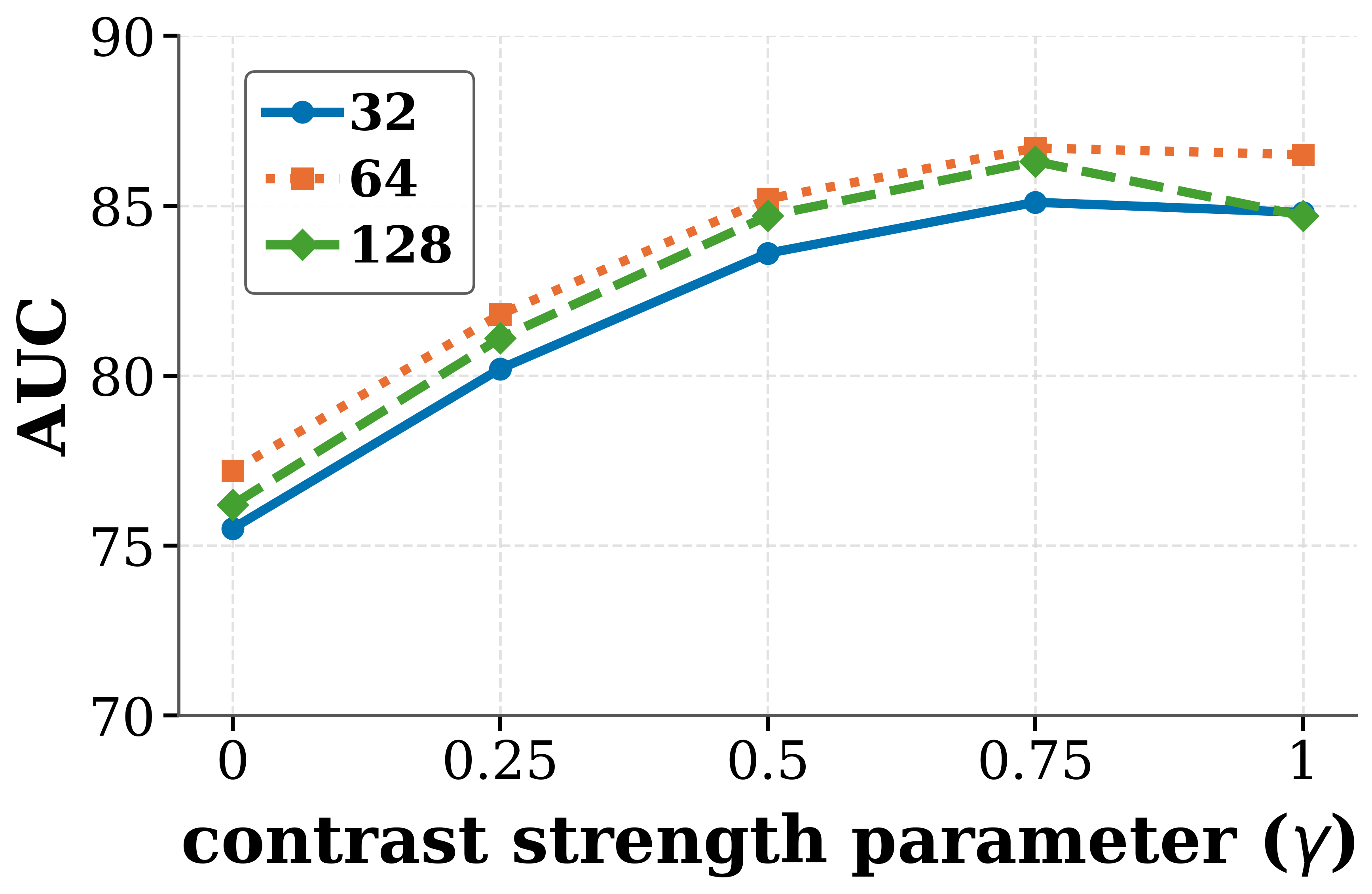}
    \caption{Results under different contrast strengths $\gamma$.}
    \label{fig:abla_gamma}
    \vspace{-10pt}
\end{figure}

\paragraph{Sensitivity to Contrast Strength.}
Following \citet{conrecall}, we analyze a generalized form of the contrastive score in \eqnref{eq:contrastive_score}:
$S_\gamma(x)=\big(\widehat{LL}_{\mathrm{nm}}(x)-\gamma\widehat{LL}_{\mathrm{m}}(x)\big)/\widehat{LL}_{0}(x)$,
where $\gamma$ controls the relative contribution of the member-prefix likelihood. 
When $\gamma=1$, this generalized score reduces to the full contrastive score used by WPMIA.
As shown in \figref{fig:abla_gamma}, setting $\gamma=0$ removes the member-prefix contrast and gives the weakest performance.
As $\gamma$ increases, the AUC improves consistently across all lengths, confirming the importance of contrasting the non-member- and member-prefix likelihoods.
The performance plateaus when $\gamma$ is between $0.75$ and $1.0$.
Although $\gamma=0.75$ is slightly better for some lengths, $\gamma=1.0$ remains highly competitive and corresponds to the full contrastive formulation.
Therefore, we set $\gamma=1.0$ as the default contrast strength in all experiments.

\paragraph{Robustness to Prefix Construction.}
We examine whether WPMIA depends on a particular construction of the prefix pools by considering label corruption, domain shift, and synthetic alternatives, as shown in \tabref{tab:prefix_robustness}. 
Light label corruption reduces AUC by only 5.0 points relative to the clean setting, whereas severe corruption and complete shuffling yield a mean AUC of 48.4. This degradation indicates that WPMIA relies on a meaningful, label-aligned contrast between the two prefix pools rather than arbitrary contextual calibration.
Correctly labeled but domain-mismatched GitHub prefixes from MIMIR achieve only an AUC of 56.5, further showing that label correctness alone is insufficient without domain compatibility. In contrast, synthetic prefixes constructed following ReCaLL~\citep{recall} and Con-ReCall~\citep{conrecall} achieve 75.3 AUC, remaining within 9.5 points of the clean setting. Thus, WPMIA is not tied to the exact held-out benchmark examples, and synthetic prefixes provide a practical alternative when verified prefixes are unavailable.

%% file: tabs/wikimia_bb.tex
\begin{table*}[t]
\centering
\resizebox{\linewidth}{!}{
\setlength{\tabcolsep}{5pt}
\begin{tabular}[t]{llcccccccccc}
\toprule
\makecell[c]{\multirow{2}{*}{\textbf{Len}}} & \makecell[c]{\multirow{2}{*}{\textbf{Method}}} & \multicolumn{2}{c}{\textbf{OPT-6.7B}} & \multicolumn{2}{c}{\textbf{Pythia-6.9B}} & \multicolumn{2}{c}{\textbf{LLaMA-13B}} & \multicolumn{2}{c}{\textbf{GPT-NeoX-20B}} & \multicolumn{2}{c}{\textbf{Average}} \\
\cmidrule(lr){3-4}\cmidrule(lr){5-6}\cmidrule(lr){7-8}\cmidrule(lr){9-10}\cmidrule(lr){11-12}
&  & AUC & TPR@5\%FPR & AUC & TPR@5\%FPR & AUC & TPR@5\%FPR & AUC & TPR@5\%FPR & AUC & TPR@5\%FPR \\
\midrule
\multirow{5}{*}{32}
& PETAL & 61.4 & 10.5 & 62.8 & 15.3 & \underline{63.2} & 11.1 & 67.6 & 21.6 & 63.8 & 14.6 \\
& SaMIA & 55.4 & 5.3 & 55.2 & 8.7 & 59.4 & 10.0 & 60.1 & 10.8 & 57.5 & 8.7\\
& SimMIA$^*$ & \underline{85.1} & \underline{28.9} & \underline{82.8} & \underline{27.4} & 60.0 & 8.9 & \underline{85.9} & \underline{34.5} & \underline{78.5} & \underline{24.9} \\
& SimMIA & 78.8 & 20.8 & 71.4 & 12.9 & 62.9 & \underline{11.6} & 76.3 & 18.7 & 72.4 & 16.0 \\
& WPMIA & \textbf{89.1} & \textbf{38.7} & \textbf{84.8} & \textbf{27.9} & \textbf{75.9} & \textbf{21.6} & \textbf{89.5} & \textbf{52.6} & \textbf{84.8} & \textbf{35.2} \\
\midrule
\multirow{5}{*}{64}
& PETAL & 59.8 & 12.7 & 60.7 & 15.5 & 62.0 & 10.4 & 65.8 & 15.1 & 62.1 & 13.4\\
& SaMIA & 65.1 & 10.4 & 61.2 & 8.8 & 63.1 & \underline{10.8} & 67.1 & 11.6 & 64.1 & 10.4 \\
& SimMIA$^*$ & \underline{85.7} & \underline{28.3} & \underline{86.2} & \textbf{39.4} & 61.8 & 10.0 & \underline{84.9} & \underline{28.7} & \underline{79.7} & \underline{26.6} \\
& SimMIA& 79.8 & 22.3 & 74.4 & 14.3 & \underline{67.1} & 6.4 & 70.4 & 15.9 & 72.9 & 14.7 \\
& WPMIA & \textbf{90.2} & \textbf{57.8} & \textbf{86.5} & \underline{35.9} & \textbf{74.3} & \textbf{23.9} & \textbf{86.9} & \textbf{45.4} & \textbf{84.5} & \textbf{40.8} \\
\midrule
\multirow{5}{*}{128}
& PETAL & 64.8 & 12.5 & 65.2 & \underline{19.2} & 64.3 & \underline{13.5} & 71.5 & 34.6 & 66.5 & 20.0\\
& SaMIA & 67.2 & 11.5 & 62.4 & 7.7 & \underline{65.2} & 7.7 & 67.5 & 16.3 & 65.6 & 10.8\\
& SimMIA$^*$ & \underline{85.5} & \underline{14.4} & \underline{82.8} & 16.3 & 61.7 & 9.6 & \underline{86.2} & \underline{40.4} & \underline{79.0} & \underline{20.1} \\
& SimMIA & 78.9 & 13.5 & 73.4 & 9.6 &  60.0 & 3.8 & 71.3 & 20.2 & 70.9 & 11.8 \\
& WPMIA & \textbf{87.2} & \textbf{49.0} & \textbf{84.7} & \textbf{39.4} & \textbf{67.4} & \textbf{15.4} & \textbf{87.5} & \textbf{48.1} & \textbf{81.7} & \textbf{38.0} \\
\bottomrule
\end{tabular}
}
\caption{Black-box results on the WikiMIA benchmark.
Complete results including gray-box baselines are provided in \tabref{tab:wikimia}.
\textbf{Bold} and \underline{underlined} values indicate the best and second-best results in each column, respectively. $^*$ denotes scoring with \eqnref{eq:simmia_hard} instead of \eqnref{eq:simmia_soft}.}
\label{tab:wikimia_bb}
\vspace{-5pt}
\end{table*}

%% file: tabs/wikimia-25-bb.tex
\begin{table*}[t]
\centering
\resizebox{\linewidth}{!}{
\setlength{\tabcolsep}{5pt}
\begin{tabular}{lcccccccccc}
\toprule
\makecell[c]{\multirow{2}{*}{\textbf{Method}}} &
\multicolumn{2}{c}{\textbf{Pythia-6.9B}} &
\multicolumn{2}{c}{\textbf{Qwen3-8B-Base}} &
\multicolumn{2}{c}{\textbf{Claude-4.5-Haiku}} &
\multicolumn{2}{c}{\textbf{Gemini-2.5-Flash}} &
\multicolumn{2}{c}{\textbf{GPT-5-Chat}} \\
\cmidrule(lr){2-3}\cmidrule(lr){4-5}\cmidrule(lr){6-7}\cmidrule(lr){8-9}\cmidrule(lr){10-11}
& AUC & TPR@5\%FPR & AUC & TPR@5\%FPR & AUC & TPR@5\%FPR & AUC & TPR@5\%FPR & AUC & TPR@5\%FPR \\
\midrule
PETAL & 72.2 & 30.1 & 62.8 & \underline{18.6} & - & - & 67.1 & 23.9 & 78.0 & 35.4\\
SaMIA  & 57.8 & 8.8 & 50.3 & 2.7 & 62.6 & \underline{8.0} & 57.0 & 15.0 & 65.7 & 13.3 \\
SimMIA$^*$ & 88.1 & 37.2 & 63.2 & 2.7 & \underline{65.7} & 6.2 & \underline{75.7} & \underline{30.2} & \underline{90.6} & \textbf{73.3}\\
SimMIA & \underline{88.8} & \underline{47.8} & \underline{63.0} & 6.2 & 60.5 & 6.2 & 74.1 & 28.5 & 83.3 & 37.2\\
WPMIA & \textbf{89.0} & \textbf{56.6} & \textbf{65.9} & \textbf{23.0} & \textbf{66.3} & \textbf{22.1} & \textbf{78.2} & \textbf{36.5} & \textbf{91.0} &  \underline{67.3}\\
\bottomrule
\end{tabular}
}
\caption{Black-box results on the WikiMIA-25 benchmark. 
Complete results including gray-box baselines are provided in \tabref{tab:wikimia25}.
\textbf{Bold} and \underline{underlined} values indicate the best and second-best results in each column, respectively. $^*$ denotes scoring with \eqnref{eq:simmia_hard} instead of \eqnref{eq:simmia_soft}.}
\label{tab:wikimia25_bb}
\vspace{-5pt}
\end{table*}

%% file: tabs/abla_overall.tex
\begin{figure*}[th]
\centering
\captionsetup[subfigure]{labelformat=parens}

\newcommand{\rowH}{3.4cm}          
\setlength{\tabcolsep}{1pt}        
\renewcommand{\arraystretch}{1.0}  
\newcommand{\labW}{0.9em}          
\newcommand{\imgW}{0.31\linewidth} 

\begin{tabular}{@{}c@{\hspace{2pt}}ccc@{}}

\parbox[c][\rowH][c]{\labW}{\centering\rotatebox{90}{AUC}} &
\begin{subfigure}[c]{\imgW}\centering
  \includegraphics[width=\linewidth,height=\rowH,keepaspectratio]{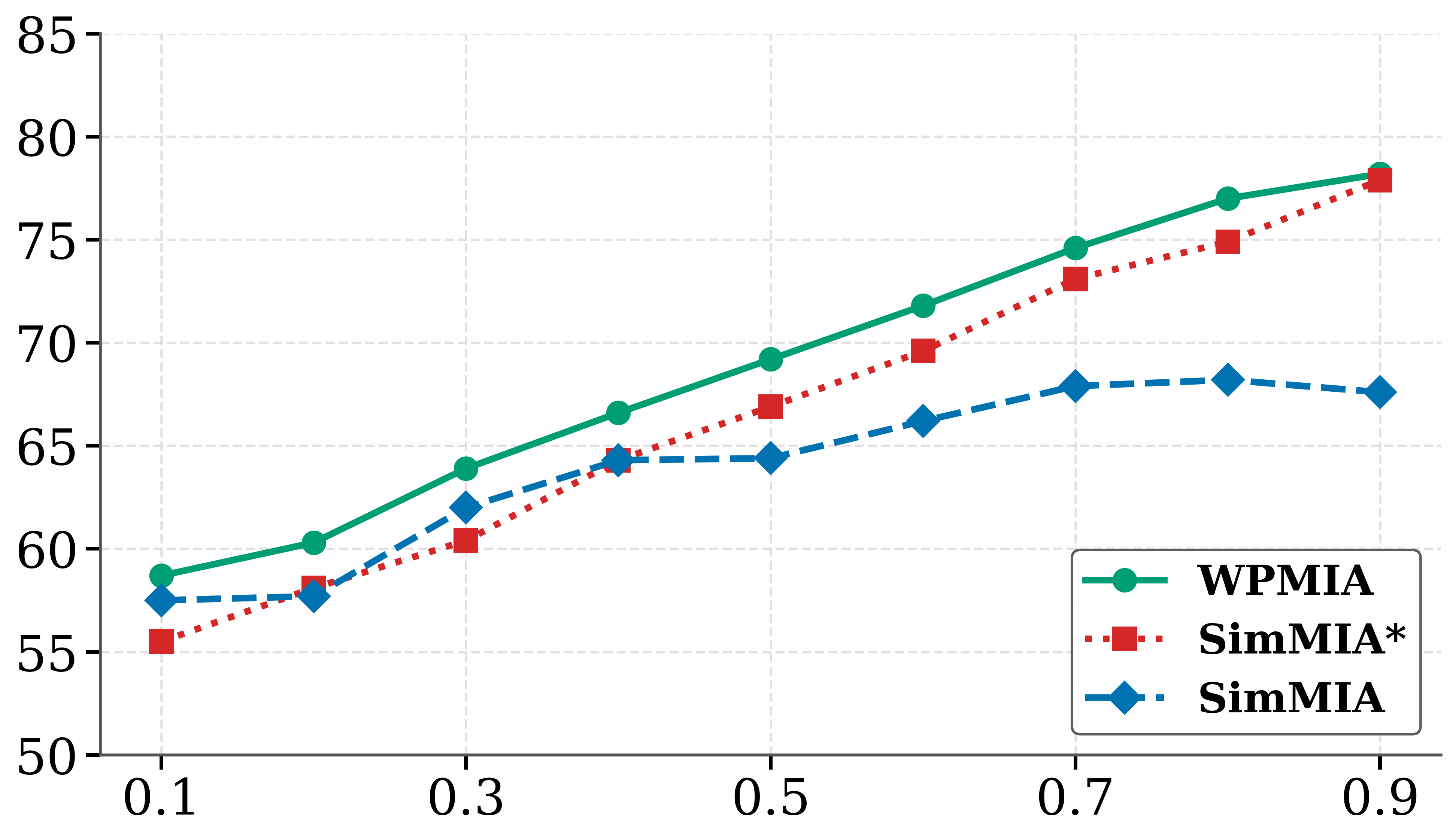}
\caption{Continuation ratios}
\label{fig:prefix-ratio} 
\end{subfigure} &
\begin{subfigure}[c]{\imgW}\centering
  \includegraphics[width=\linewidth,height=\rowH,keepaspectratio]{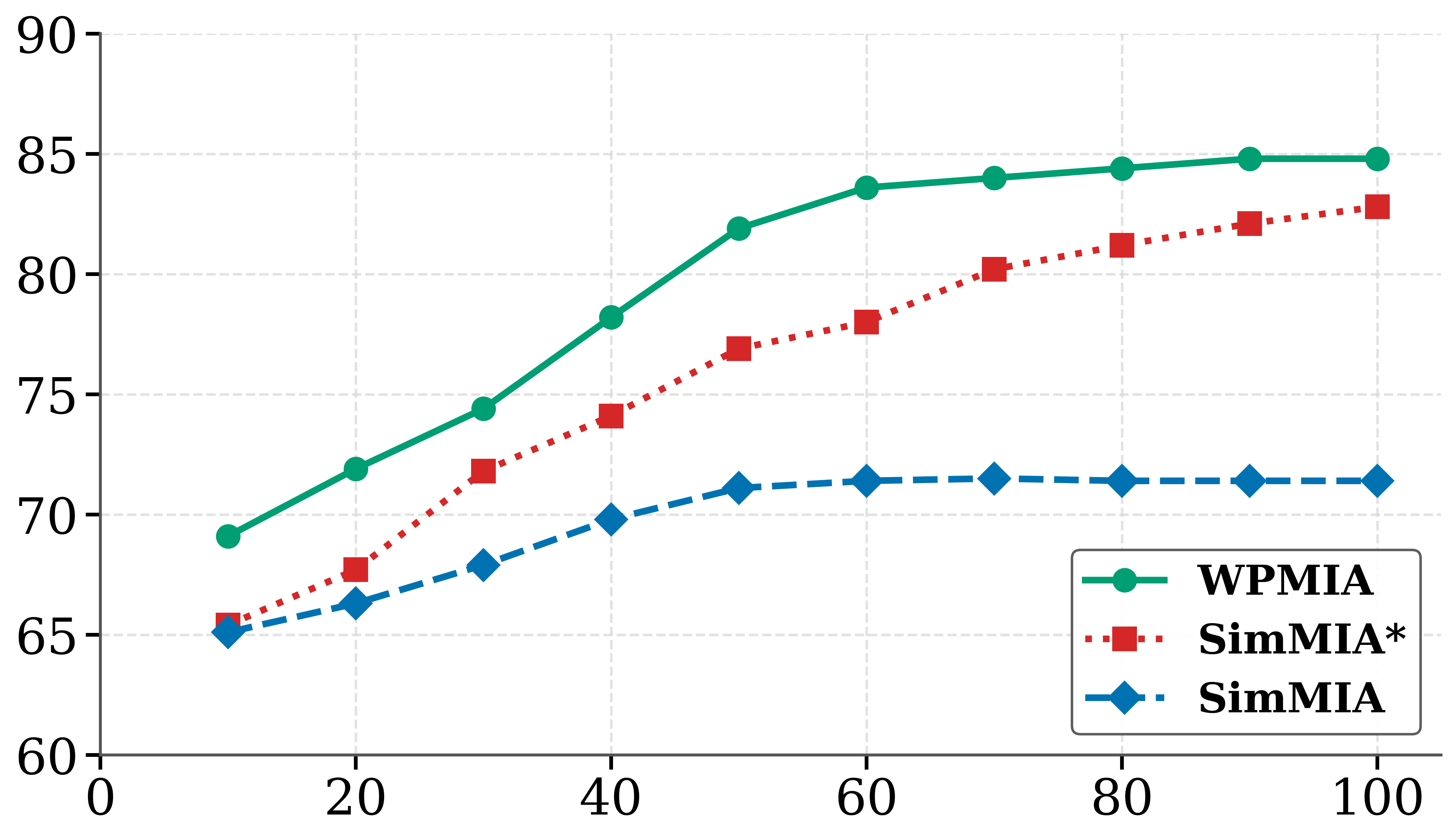}
  \caption{Sample sizes}
\label{fig:num-shots} 
\end{subfigure} &
\begin{subfigure}[c]{\imgW}\centering
  \includegraphics[width=\linewidth,height=\rowH,keepaspectratio]{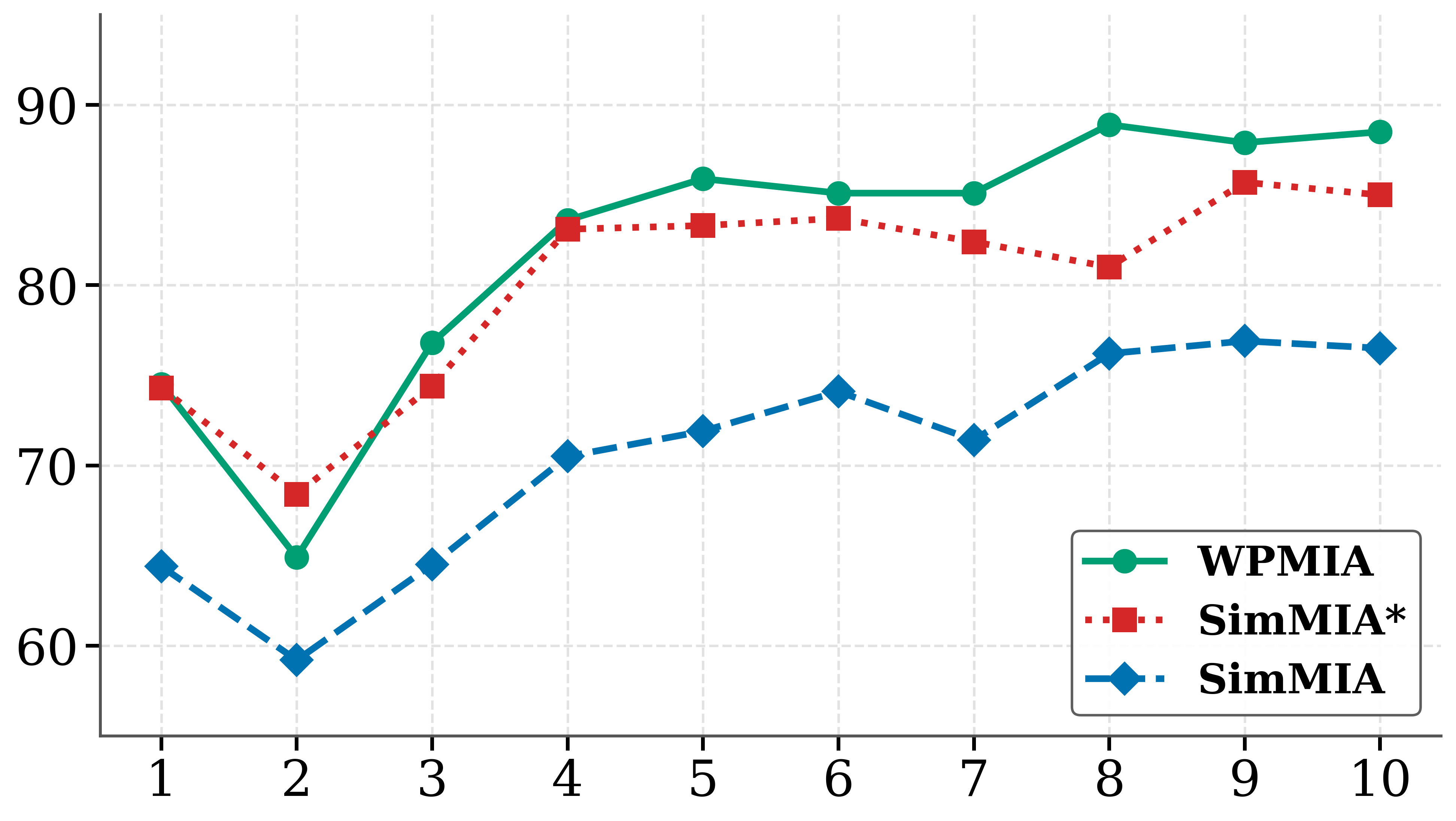}
  \caption{Prefix Shots}
\label{fig:num-samples} 
\end{subfigure}
\end{tabular}
\caption{Attack performance under different continuation ratios, sample sizes, and prefix shots at Len. 32.}
\label{fig:ablation_overall}
\vspace{-5pt}
\end{figure*}

%% file: tabs/abla_component_analysis.tex
\begin{table}[ht]
\centering
\resizebox{\linewidth}{!}{
\begin{tabular}{lccc}
\toprule
\textbf{Method} & \textbf{Len. 32} & \textbf{Len. 64} & \textbf{Len. 128} \\
\midrule
LL ($\mathcal{S}_{0}$) & 58.8 & 55.4 & 62.0 \\
\midrule
+NP, sequence ($\mathcal{S}_{nm}$) & 75.5 & 77.2 & 76.2 \\
+NP, word & 52.0 & 46.9 & 54.7 \\
\midrule
+MP, word & 51.4 & 48.8 & 48.8 \\
+MP, sequence ($\mathcal{S}_{\mathrm{ctr}}$) & \textbf{84.8} & \textbf{86.5} & \textbf{84.7} \\
\bottomrule
\end{tabular}
}
\caption{Component-wise ablation of WPMIA.}
\label{tab:wikimia-pythia-component-wise}
\end{table}

%% file: tabs/abla_estimator.tex
\begin{table}[htbp]
\centering
\resizebox{\linewidth}{!}{
\begin{tabular}{llccc}
\toprule
\textbf{Estimator} & \textbf{Prefix}
& \textbf{Len. 32} & \textbf{Len. 64} & \textbf{Len. 128} \\
\midrule

\multirow{3}{*}{Exact match}
& No prefix
& 59.87 & 58.01 & 65.08 \\
& +NP
& 73.88 & 72.63 & 68.81 \\
& +MP
& 81.00 & 80.58 & 75.50 \\

\midrule

\multirow{3}{*}{Raw similarity}
& No prefix
& 57.93 & 54.86 & 59.86 \\
& +NP
& 70.57 & 72.91 & 72.21 \\
& +MP
& 79.80 & 82.65 & 81.12 \\

\midrule

\multirow{3}{*}{Semantic kernel}
& No prefix
& 58.75 & 55.40 & 61.99 \\
& +NP
& 75.47 & 77.19 & 76.16 \\
& +MP
& \textbf{84.78} & \textbf{86.48} & \textbf{84.67} \\

\bottomrule
\end{tabular}
}
\caption{Crossed ablation of word-level estimators and prefix-conditioning strategies. \textit{+NP} denotes non-member-prefix conditioning, whereas \textit{+MP} additionally introduces member-prefix contrast.}
\label{tab:wikimia_pythia_estimator_ablation}
\end{table}

%% file: tabs/abla_embedding.tex
\begin{table}[ht]
\centering
\resizebox{\linewidth}{!}{
\setlength{\tabcolsep}{3pt}
\renewcommand{\arraystretch}{1.08}
\begin{tabular}{l|ccc}
\toprule
\textbf{Embedding Model} & \textbf{Len. 32} & \textbf{Len. 64} & \textbf{Len. 128} \\
\midrule
Word2Vec & 61.5 & 58.2 & 62.2 \\
fastText & 80.9 & 82.1 & 82.2 \\
\midrule
bge-large-en-v1.5 & 86.3 & 88.0 & 84.9 \\
all-MiniLM-L6-v2 & 84.8 & 86.5 & 84.7 \\
UAE-Large-V1 & \textbf{86.5} & \textbf{88.3} & \textbf{85.1} \\
mxbai-embed-large-v1 & 86.3 & 88.0 & 84.8 \\
\bottomrule
\end{tabular}
}
\caption{Results with different embedding models.}
\label{tab:wikimia_embedding_ablation}
\end{table}

%% file: tabs/abla_robustness_prefix.tex
\begin{table*}[thbp]
\centering
\resizebox{\linewidth}{!}{
\begin{tabular}{llcrr}
\toprule
\textbf{Prefix condition}
& \textbf{Construction}
& \textbf{Mislabeled prefixes}
& \textbf{AUC}
& \textbf{TPR@5\%FPR} \\
\midrule
Clean
& Correct WikiMIA member/non-member prefix pools
& 0/14 & 84.78 & 27.89 \\
\midrule
Noisy-light
& Swap one prefix from each pool
& 2/14 & 79.76 & 15.79 \\
\midrule
Noisy-heavy
& Swap three prefixes from each pool
& 6/14 & 48.73 & 6.05 \\
\midrule
Fully shuffled
& Randomly repartition all prefixes with a fixed score direction
& 8/14 & 48.14 & 5.08 \\
\midrule
GitHub OOD
& Correctly labeled GitHub subset prefixes from MIMIR
& N/A & 56.54 & 8.11 \\
\midrule
Synthetic
& Synthetic non-member and member-like prefixes
& N/A & 75.27 & 14.84 \\
\bottomrule
\end{tabular}
}
\caption{Robustness to prefix-pool corruption and distribution shift at Len.~32. Each condition uses seven member and seven non-member prefixes.}
\label{tab:prefix_robustness}
\vspace{-5pt}
\end{table*}

%% file: sec/4_conclusion.tex
\section{Conclusion}\label{sec:conclusion}

This work presents WPMIA, a strict black-box membership inference attack for auditing the privacy risks of LLMs. 
Unlike existing gray-box MIAs that require access to the target tokenizer and per-token logits, WPMIA recovers likelihood-based membership signals directly from textual continuations. 
It estimates word-level generation probabilities through Monte Carlo sampling with semantic kernel smoothing, aggregates these estimates into sequence-level likelihoods, and further strengthens the membership signal through contrastive prefix conditioning. 
Extensive experiments on WikiMIA, MIMIR, and WikiMIA-25 show that WPMIA achieves the strongest black-box performance on open-source LLMs and remains effective on recent proprietary LLMs, demonstrating the practical feasibility of strict black-box privacy auditing. 

\section*{Limitations}\label{sec:limitation}

Despite its effectiveness, WPMIA has a limitation common to sampling-based black-box MIAs for LLMs: it requires repeated sampling to recover membership signals.
Unlike gray-box methods, which can obtain sequence-level likelihoods with a single forward pass, WPMIA must repeatedly query the target model and estimate word-level probabilities from sampled textual continuations.
In addition, its contrastive prefix mechanism requires estimating conditional likelihoods under both member and non-member prefixes, thereby increasing the overall query budget.
This cost becomes more substantial for long target texts and proprietary LLM APIs, where repeated queries incur additional computation, latency, and monetary expense.
An important direction for future work is therefore to improve the sample efficiency of WPMIA, such as through adaptive sampling, variance-reduced estimation, or early-stopping strategies that reduce the number of required continuations while preserving reliable detection performance.

Another limitation is that our evaluation is restricted to English text. 
The effectiveness of WPMIA in multilingual settings may be affected by language-specific tokenization, semantic embeddings, and prefix construction.
Future work should therefore evaluate WPMIA on multilingual benchmarks and investigate language-specific adaptations.

\section*{Ethical Considerations}

This work studies membership inference attacks for privacy auditing of LLMs and is intended to help researchers and model providers assess potential training-data leakage under strict black-box access.
However, membership inference is inherently dual-use and could be misused to probe for the presence of sensitive or private texts in training data; therefore, WPMIA should be applied only in authorized auditing, controlled research, and privacy-risk evaluation settings.

%% file: sec/X_appendix.tex
\appendix

\section{Dataset Statistics}\label{app:dataset}

Detailed statistics of WikiMIA, MIMIR, and WikiMIA-25 are summarized in \tabrefs{tab:wikimia_dataset_statistics}{tab:wikimia25_dataset_statistics}.

\begin{table}[h]
\centering
\small
\begin{tabular}{lccc}
\toprule
\multirow{2}{*}{\textbf{Statistic}} 
& \multicolumn{3}{c}{\textbf{Text Length}} \\
\cmidrule(lr){2-4}
& \textbf{32} & \textbf{64} & \textbf{128} \\
\midrule
Total examples      & 776    & 542    & 250    \\
Non-member Ratio   & 50.1\% & 47.6\% & 44.4\% \\
Member Ratio       & 49.9\% & 52.4\% & 55.6\% \\
\bottomrule
\end{tabular}
\caption{WikiMIA dataset statistics.}
\label{tab:wikimia_dataset_statistics}
\end{table}

\begin{table}[h]
\centering
\small
\begin{tabular}{lc}
\toprule
\textbf{Subset} & \textbf{Samples} \\
\midrule
wikipedia\_(en)  & 2000 \\
github           & 536  \\
pile\_cc         & 2000 \\
pubmed\_central  & 982  \\
arxiv            & 1000 \\
dm\_mathematics  & 178  \\
hackernews       & 1292 \\
\bottomrule
\end{tabular}
\caption{MIMIR dataset statistics under the 7-gram setting.}
\label{tab:mimir_dataset_statistics}
\end{table}

\begin{table}[h]
\centering
\resizebox{\linewidth}{!}{
\begin{tabular}{lccc}
\toprule
\textbf{Text Length} & \textbf{Members} & \textbf{Non-members} & \textbf{Total Samples} \\
\midrule
32 & 120 & 120 & 240 \\
\bottomrule
\end{tabular}
}
\caption{WikiMIA-25 subset used for cost-efficient evaluation.}
\label{tab:wikimia25_dataset_statistics}
\end{table}

\section{Additional Implementation Details}\label{app:implementation}

\begin{table}[htbp]
\centering
\resizebox{\linewidth}{!}{
\begin{tabular}{lccc}
\toprule
\textbf{Setting} & \textbf{WikiMIA} & \textbf{MIMIR} & \textbf{WikiMIA-25} \\
\midrule
Monte Carlo samples $M$ & 100 & 100 & 10 \\
Prefix shots $T$ & 7 & 10 & 7 \\
Smoothing constant $\epsilon$ & $10^{-8}$ & $10^{-8}$ & $10^{-8}$ \\
Kernel temperature $\tau$ & 0.2 & 0.2 & 0.1 \\
\bottomrule
\end{tabular}
}
\caption{Hyperparameter settings used for WPMIA.}
\label{tab:wpmia_hyperparameters}
\end{table}

\paragraph{Benchmark Prefix Construction.}
WPMIA uses both member-prefix and non-member-prefix conditioning.
For each benchmark, we reserve an equal-sized pool of member and non-member examples for prefix construction.
The reserved prefix examples are removed from the evaluation set for all methods, including baselines, to ensure a fair comparison.
For WikiMIA and WikiMIA-25, we use $T=7$ prefix shots.
For MIMIR, we use $T=10$ prefix shots.

\paragraph{Hyperparameter Settings.} 
The hyperparameter settings used for WPMIA across different benchmarks are summarized in \tabref{tab:wpmia_hyperparameters}.

\paragraph{Prompt Template for Proprietary LLMs.}
For proprietary LLMs, we use the following system prompt based on SimMIA~\cite{simmia}. The prompt constrains the model to generate a single next word or punctuation mark for a given \texttt{[PREFIX]}.

\begin{center}
    \begin{tabular}{|p{0.85\linewidth}|}
    \hline
    \textbf{Prompt} \\
    \hline
    \small
    \textbf{System:} Return ONLY the single next token from the text. It can be punctuation or one whole word. No spaces, no quotes, no extra text.
    
    \medskip
    \textbf{User:} Text so far: \texttt{[PREFIX]} \\
    \hline
    \end{tabular}
\end{center}

\section{Practical Prefix Construction}

We next discuss how to instantiate the benchmark prefix pools in realistic black-box auditing settings. Depending on the auditor's access to training-data information, prefix construction can range from verified member/non-member examples to high-confidence or fully synthetic proxies.

\paragraph{Authorized audits with verified prefixes.}
In authorized audits, a model provider or data curator may have access to examples whose membership status is known. Member prefixes can be drawn directly from the training corpus or from pre-inserted canary documents whose inclusion in training is verifiable~\cite{sharer}. Non-member prefixes can be constructed from newly created private documents that were never exposed to the model provider. This setting most closely matches the benchmark construction, where the two prefix pools have reliable membership labels.

\paragraph{External audits with high-confidence proxies.}
When direct access to the training set is unavailable, an external auditor can instead construct prefix pools using high-confidence membership proxies. For a fixed model snapshot with a known training cutoff, newly created private documents or public documents released after the cutoff can serve as non-member candidates. 
Member-side proxies can be drawn from pre-cutoff public corpora whose likely inclusion is supported by model documentation or established pretraining practice. For example, Wikipedia, GitHub, and arXiv are standard components of the Pile~\cite{pile}, which was used to train the Pythia model family~\cite{pythia}. Such examples do not provide verified membership labels, but they offer a practical approximation when the model's training sources and cutoff are partially known.

\paragraph{External audits with synthetic proxies.}
When neither verified examples nor sufficiently reliable public proxies are available, both prefix pools can be constructed synthetically.
ReCaLL~\citep{recall} uses GPT-4o to generate passages that match the length, domain, and style of the target text, which can serve as approximate non-member prefixes. Con-ReCall~\citep{conrecall} further introduces a procedure for constructing member-like prefixes without access to the original training set. It first uses GPT-4o to generate descriptions of significant events, partially truncates these descriptions, and then prompts the target model to complete them. The resulting target-model continuations are used as member-like prefixes because they reflect knowledge and language patterns internalized by the target model. Using these synthetic prefixes also achieves appreciable attack performance, as reported in \tabref{tab:prefix_robustness}.

\section{Additional Experiments}\label{app:results}

\paragraph{Complete Benchmark Results.}
\tabrefs{tab:wikimia}{tab:wikimia25} report the complete comparisons on WikiMIA,
MIMIR, and WikiMIA-25, respectively.
These results complement the black-box comparisons in the main manuscript with the corresponding gray-box baselines. Across the three benchmarks, WPMIA is the strongest black-box method in most settings and remains competitive with gray-box.

\paragraph{Budget-Matched Comparison.}
Beyond comparing sampling-based black-box attacks with the same sample size, we further compare them under approximately matched budgets of generated tokens.
Generated output tokens provide a common cost measure that accounts for the long continuations used by SaMIA and the short first-word outputs used by WPMIA.
We compare SaMIA, SimMIA, and WPMIA at low, medium, and high budgets, using SimMIA as the cost reference with $M=30$, $M=60$, and $M=100$, respectively. At each level, we select the integer sample sizes for SaMIA and WPMIA whose generated-token counts are closest to the corresponding SimMIA budget.
As shown in \tabref{tab:wikimia-pythia-token-normalized}, WPMIA achieves the best performance across all budget levels.
Averaged across the three budgets, it achieves an AUC of 83.1 and a TPR@5\%FPR of 27.9, exceeding SimMIA by 12.4 and 15.4 points, respectively. 
These results show that WPMIA remains the strongest method when the attacks are compared under approximately matched generated-token budgets.

\input{tabs/abla_token_normalized}

\paragraph{Proprietary API Statistics.}
We report the request count, runtime, and monetary cost of the proprietary LLM experiments in \tabref{tab:proprietary_api_cost}.
For a record with $L$ scored word positions, WPMIA requires approximately $Q=3ML$ requests because each position is evaluated under the unprefixed, member-prefix, and non-member-prefix conditions.
WikiMIA-25 contains 7,898 scored positions in total, resulting in an expected retry-free total of 236,940 requests when $M=10$.
The observed HTTP-response counts in \tabref{tab:proprietary_api_cost} are slightly higher than this estimate because interrupted or failed requests may trigger retries.
Provider-specific throughput and pricing lead to different practical trade-offs: Claude-4.5-Haiku achieves the lowest mean runtime per record, whereas Gemini-2.5-Flash incurs the lowest total monetary cost.

\input{tabs/api_cost}

\paragraph{Stability across Sample Sizes.}
We repeat the evaluation of Pythia-6.9B on WikiMIA using five sampling seeds to assess the stability of WPMIA across different sample sizes.
As shown in \tabref{tab:sampling_stability}, WPMIA exhibits moderate variability even at $M=10$, with AUC standard deviations of 1.10, 2.77, and 1.44 for lengths 32, 64, and 128, respectively.
Increasing $M$ consistently improves the mean AUC across all three lengths, while the marginal gains diminish around $M=70$--$80$, particularly for lengths 32 and 64. These results reveal a clear trade-off between sampling cost and attack performance.

\input{tabs/abla_sample_size_stability}

\paragraph{Fine-Grained Temperature Sensitivity.}
We further examine temperature sensitivity by sweeping $\tau$ from 0.10 to 0.50 in increments of 0.05.
As shown in \tabref{tab:temperature_fine}, the AUC averaged across sequence lengths ranges only from 83.6 to 85.4 over the full sweep. The variation is even smaller within $\tau\in[0.15,0.30]$, where the average AUC remains between 84.6 and 85.4.
The default value $\tau=0.2$ achieves an average AUC of 85.3, within 0.1 points of the best tested setting. These results indicate that WPMIA is stable over a broad temperature range on WikiMIA and support using $\tau=0.2$ uniformly for this benchmark.
We apply the same value across the MIMIR domains without additional tuning.

\input{tabs/abla_temperature_finegrained}

\section{Pseudocode}\label{app:pseudo-code}

We present the pipeline of WPMIA in Algorithm~\ref{alg:wpmia}.

\input{tabs/pseudocode}

\input{tabs/wikimia}

\input{tabs/mimir}

\input{tabs/wikimia-25}

%% file: tabs/abla_token_normalized.tex
\begin{table}[ht]
\centering
\resizebox{\linewidth}{!}{
\begin{tabular}{llcrrr}
\toprule
\multirow{2}{*}{\textbf{Budget}}
& \multirow{2}{*}{\textbf{Method}}
& \multirow{2}{*}{$\boldsymbol{M}$}
& \multicolumn{1}{c}{\textbf{Avg. generated}}
& \multirow{2}{*}{\textbf{AUC}}
& \multirow{2}{*}{\textbf{TPR@5\%FPR}} \\
& & &
\multicolumn{1}{c}{\textbf{tokens}}
& & \\
\midrule
\multirow{3}{*}{Low}
& SaMIA  & 10 & 6,317.68  & 51.30          & 5.53 \\
& SimMIA & 30 & 6,313.98  & 69.99          & 11.58 \\
& WPMIA  & 20 & 6,313.86  & \textbf{81.25} & \textbf{28.68} \\

\midrule

\multirow{3}{*}{Medium}
& SaMIA  & 20 & 12,499.20 & 51.42          & 6.58 \\
& SimMIA & 60 & 12,627.62 & 70.88          & 13.16 \\
& WPMIA  & 40 & 12,628.24 & \textbf{83.40} & \textbf{27.89} \\

\midrule

\multirow{3}{*}{High}
& SaMIA  & 34  & 21,106.13 & 50.69          & 8.16 \\
& SimMIA & 100 & 21,045.00 & 71.24          & 12.63 \\
& WPMIA  & 67  & 21,152.33 & \textbf{84.59} & \textbf{27.11} \\

\bottomrule
\end{tabular}
}
\caption{Generated-token-budget-matched comparison of sampling-based
black-box attacks on WikiMIA with Pythia-6.9B at length 32.}
\label{tab:wikimia-pythia-token-normalized}
\end{table}

%% file: tabs/api_cost.tex
\begin{table}[htbp]
\centering
\resizebox{\linewidth}{!}{
\begin{tabular}{lrrr}
\toprule
\multirow{2}{*}{\textbf{Model}}
& \multicolumn{1}{c}{\textbf{HTTP}}
& \multicolumn{1}{c}{\textbf{Runtime/}}
& \multicolumn{1}{c}{\textbf{Monetary cost}} \\
& \multicolumn{1}{c}{\textbf{responses}}
& \multicolumn{1}{c}{\textbf{record (s)}}
& \multicolumn{1}{c}{\textbf{(USD)}} \\
\midrule
GPT-5-Chat
& 241,492 & 216.43 & \$266.8 \\
\midrule
Gemini-2.5-Flash
& 238,649 & 119.81 & \$78.0 \\
\midrule
Claude-Haiku-4.5
& 239,293 & 86.25 & \$240.7 \\
\bottomrule
\end{tabular}
}
\caption{API statistics for WPMIA on WikiMIA-25 with $M=10$. HTTP response counts include retries; runtime is averaged per record at a concurrency level of 32, and monetary cost is reported in U.S. dollars.}
\label{tab:proprietary_api_cost}
\end{table}

%% file: tabs/abla_sample_size_stability.tex
\begin{table}[htbp]
\centering
\resizebox{\linewidth}{!}{
\begin{tabular}{lccc}
\toprule
$M$ & Len. 32 & Len. 64 & Len. 128 \\
\midrule
10  & $68.32\pm1.10$ & $71.74\pm2.77$ & $67.63\pm1.44$ \\
20  & $71.84\pm0.96$ & $75.49\pm1.82$ & $72.12\pm2.30$ \\
30  & $75.11\pm0.83$ & $79.00\pm0.84$ & $77.80\pm1.75$ \\
40  & $78.32\pm0.49$ & $80.74\pm0.72$ & $80.58\pm2.53$ \\
50  & $80.90\pm0.64$ & $83.04\pm0.70$ & $83.30\pm2.12$ \\
60  & $82.31\pm0.83$ & $84.57\pm0.72$ & $84.66\pm1.47$ \\
70  & $83.17\pm0.49$ & $85.77\pm0.35$ & $86.26\pm1.51$ \\
80  & $83.58\pm0.55$ & $86.10\pm0.22$ & $87.33\pm1.80$ \\
90  & $84.13\pm0.41$ & $86.41\pm0.22$ & $87.95\pm1.88$ \\
100 & $84.48\pm0.29$ & $86.67\pm0.24$ & $87.99\pm1.88$ \\
\bottomrule
\end{tabular}
}
\caption{Mean AUC ($\pm$ standard deviation) over five sampling seeds for WPMIA on WikiMIA with Pythia-6.9B, across sample sizes $M$ and sequence lengths.}
\label{tab:sampling_stability}
\end{table}

%% file: tabs/abla_temperature_finegrained.tex
\begin{table}[htbp]
\centering
\begin{tabular}{lrrr}
\toprule
$\tau$ & Len. 32 & Len. 64 & Len. 128 \\
\midrule
0.10 & 84.69 & 85.00 & 81.56 \\
0.15 & \textbf{85.30} & \textbf{86.52} & 84.26 \\
0.20 & 84.78 & 86.48 & 84.67 \\
0.25 & 84.16 & 86.17 & \textbf{84.74} \\
0.30 & 83.63 & 85.84 & 84.44 \\
0.35 & 83.21 & 85.58 & 84.17 \\
0.40 & 82.87 & 85.29 & 83.90 \\
0.45 & 82.60 & 85.10 & 83.69 \\
0.50 & 82.33 & 84.93 & 83.50 \\
\bottomrule
\end{tabular}
\caption{AUC of WPMIA on WikiMIA with Pythia-6.9B under different
semantic-kernel temperatures.}
\label{tab:temperature_fine}
\end{table}

%% file: tabs/pseudocode.tex
\begin{algorithm*}[t]
\caption{Word-level Probability Membership Inference Attack}
\label{alg:wpmia}
\begin{algorithmic}[1]
\State \textbf{Input:} target text $x=(x_1,x_2,\ldots,x_L)$, language model $f_\theta$, number of samples $M$, 
non-member prefix pool $\mathcal{P}_{\mathrm{nm}}$, member prefix pool $\mathcal{P}_{\mathrm{m}}$, 
number of prefix shots $T$, kernel function $K_{\tau}$, smoothing constant $\epsilon$, decision threshold $\kappa$
\State \textbf{Output:} membership score $\mathcal{S}(\theta,x)$ and decision on whether $x$ is included in the training data of $f_\theta$ $(1$ or $0)$
\State $P_0 \gets \langle\rangle$
\State $P_{\mathrm{nm}} \gets p^{\mathrm{nm}}_1 \oplus p^{\mathrm{nm}}_2 \oplus \cdots \oplus p^{\mathrm{nm}}_T$, where $p^{\mathrm{nm}}_t \in \mathcal{P}_{\mathrm{nm}}$
\State $P_{\mathrm{m}} \gets p^{\mathrm{m}}_1 \oplus p^{\mathrm{m}}_2 \oplus \cdots \oplus p^{\mathrm{m}}_T$, where $p^{\mathrm{m}}_t \in \mathcal{P}_{\mathrm{m}}$
\For{$i=1$ \textbf{to} $L$}
    \For{$q \in \{0,\mathrm{nm},\mathrm{m}\}$}
        \State $c_i^q \gets P_q \oplus (x_1,x_2,\ldots,x_{i-1})$
        \For{$j=1$ \textbf{to} $M$}
            \State $\widehat{x}^{c_i^q}_{i,j} \gets \textsc{FirstWord}\!\left(f_\theta(c_i^q)\right)$
            \Comment{sample the first generated word from $f_\theta$ using $c_i^q$ as the prompt}
        \EndFor
        \State $\widehat{p}(x_i|c_i^q) \gets \frac{1}{M}\sum_{j=1}^{M} K_{\tau}\!\left(x_i,\widehat{x}^{c_i^q}_{i,j}\right)$
        \Comment{estimate word-level probability mass}
        \State $\widehat{\ell}(x_i|c_i^q) \gets \log\!\left(\epsilon+\widehat{p}(x_i|c_i^q)\right)$
    \EndFor
\EndFor
\For{$q \in \{0,\mathrm{nm},\mathrm{m}\}$}
    \State $\widehat{LL}^{q}(x) \gets \frac{1}{L}\sum_{i=1}^{L} \widehat{\ell}(x_i|c_i^q)$
    \Comment{aggregate word-level log-probabilities into sequence-level log-likelihood}
\EndFor
\State $\mathcal{S}(\theta,x) \gets \frac{\widehat{LL}^{\mathrm{nm}}(x)-  \widehat{LL}^{\mathrm{m}}(x)}{\widehat{LL}^{0}(x)}$
\Comment{contrast prefix-conditioned log-likelihoods}
\If{$\mathcal{S}(\theta,x) > \kappa$}
    \State \Return $\mathcal{S}(\theta,x), 1$
\Else
    \State \Return $\mathcal{S}(\theta,x), 0$
\EndIf
\end{algorithmic}
\end{algorithm*}

%% file: tabs/wikimia.tex
\begin{table*}[t!]
\centering
\resizebox{\linewidth}{!}{
\setlength{\tabcolsep}{5pt}
\begin{tabular}[t]{llcccccccccc}
\toprule
\makecell[c]{\multirow{2}{*}{\textbf{Len}}} & \makecell[c]{\multirow{2}{*}{\textbf{Method}}} & \multicolumn{2}{c}{\textbf{OPT-6.7B}} & \multicolumn{2}{c}{\textbf{Pythia-6.9B}} & \multicolumn{2}{c}{\textbf{LLaMA-13B}} & \multicolumn{2}{c}{\textbf{GPT-NeoX-20B}} & \multicolumn{2}{c}{\textbf{Average}} \\
\cmidrule(lr){3-4}\cmidrule(lr){5-6}\cmidrule(lr){7-8}\cmidrule(lr){9-10}\cmidrule(lr){11-12}
&  & AUC & TPR@5\%FPR & AUC & TPR@5\%FPR & AUC & TPR@5\%FPR & AUC & TPR@5\%FPR & AUC & TPR@5\%FPR \\
\midrule
\multirow{14}{*}{32} & \multicolumn{11}{c}{\cellcolor{gray!30}\texttt{Gray-Box}} \\
& Loss & 60.3 & 10.5 & 63.5 & 13.7 & 67.4 & 13.7 & 68.7 & 19.0 & 65.0 & 14.2\\
& Reference & 64.0 & 7.9 & 63.8 & 6.3 & 58.2 & 4.7 & 67.9 & 16.6 & 63.5 & 8.9\\
& Lowercase & 57.7 & 8.9 & 58.9 & 11.8 & 65.3 & 13.4 & 61.6 & 14.5 & 60.9 & 12.2 \\
& Zlib & 61.3 & 12.6 & 64.1 & 16.8 & 67.7 & 11.8 & 69.0 & 20.8 & 65.5 & 15.5\\
& Neighbor & 64.5 & 12.1 & 65.7 & \underline{17.1} & 66.2 & 11.8 & 70.4 & 22.9 & 66.7 & 16.0\\
& Min-K\% & 61.1 & \underline{16.3} & 69.5 & 13.2 & 66.5 & 18.4 & 71.5 & \underline{28.2} & 67.2 & \underline{19.0} \\
& Min-K\%++ & \underline{65.1} & 8.9 & \underline{70.0} & 13.9 & \underline{84.5} & \underline{32.4} & \underline{74.2} & 19.0 & \underline{73.5} & 18.6\\
& ReCaLL & \textbf{77.6} & \textbf{25.5} & \textbf{86.1} & \textbf{28.9} & \textbf{89.1} & \textbf{41.8} & \textbf{87.9} & \textbf{34.7} & \textbf{85.2} & \textbf{32.7}\\
\cmidrule{2-12}
& \multicolumn{11}{c}{\cellcolor{gray!30}\texttt{Black-Box}} \\
& PETAL & 61.4 & 10.5 & 62.8 & 15.3 & \underline{63.2} & 11.1 & 67.6 & 21.6 & 63.8 & 14.6 \\
& SaMIA & 55.4 & 5.3 & 55.2 & 8.7 & 59.4 & 10.0 & 60.1 & 10.8 & 57.5 & 8.7\\
& SimMIA$^*$ & \underline{85.1} & \underline{28.9} & \underline{82.8} & \underline{27.4} & 60.0 & 8.9 & \underline{85.9} & \underline{34.5} & \underline{78.5} & \underline{24.9} \\
& SimMIA & 78.8 & 20.8 & 71.4 & 12.9 & 62.9 & \underline{11.6} & 76.3 & 18.7 & 72.4 & 16.0 \\
& WPMIA & \textbf{89.1} & \textbf{38.7} & \textbf{84.8} & \textbf{27.9} & \textbf{75.9} & \textbf{21.6} & \textbf{89.5} & \textbf{52.6} & \textbf{84.8} & \textbf{35.2} \\
\midrule
\multirow{14}{*}{64} & \multicolumn{11}{c}{\cellcolor{gray!30}\texttt{Gray-Box}} \\
& Loss & 56.5 & 13.9 & 60.0 & 13.1 & 63.4 & 12.0 & 65.8 & 12.0 & 61.4 & 12.8\\
& Reference & 62.5 & 5.2 & 63.1 & 12.4 & 63.8 & 4.0 & 66.2 & 15.9 & 63.9 & 9.4 \\
& Lowercase & 55.6 & 11.2 & 58.7 & 8.4 & 61.8 & 8.4 & 60.4 & 12.4 & 59.1 & 10.1\\
& Zlib & 59.2 & 12.0 & 62.0 & 15.9 & 65.2 & 12.7 & 67.5 & 17.1 & 63.5 & 14.4\\
& Neighbor & 60.7 & 12.0 & 63.6 & 9.6 & 64.3 & 9.6 & 67.9 & 12.4 & 64.1 & 10.9\\
& Min-K\% & 59.8 & 16.7 & 64.0 & 20.3 & 65.9 & 18.3 & 71.7 & 18.7 & 65.4 & 18.5\\
& Min-K\%++ & \underline{64.9} & \underline{17.1} & \underline{71.7} & \underline{21.9} & \underline{84.6} & \underline{29.5} & \underline{77.2} & \underline{25.5} & \underline{74.6} & \underline{23.5}\\
& ReCaLL & \textbf{76.3} & \textbf{23.9} & \textbf{89.8} & \textbf{49.8} & \textbf{89.5} & \textbf{48.6} & \textbf{87.1} & \textbf{33.5} & \textbf{85.7} & \textbf{39.0}\\
\cmidrule{2-12}
& \multicolumn{11}{c}{\cellcolor{gray!30}\texttt{Black-Box}} \\
& PETAL & 59.8 & 12.7 & 60.7 & 15.5 & 62.0 & 10.4 & 65.8 & 15.1 & 62.1 & 13.4\\
& SaMIA & 65.1 & 10.4 & 61.2 & 8.8 & 63.1 & \underline{10.8} & 67.1 & 11.6 & 64.1 & 10.4 \\
& SimMIA$^*$ & \underline{85.7} & \underline{28.3} & \underline{86.2} & \textbf{39.4} & 61.8 & 10.0 & \underline{84.9} & \underline{28.7} & \underline{79.7} & \underline{26.6} \\
& SimMIA& 79.8 & 22.3 & 74.4 & 14.3 & \underline{67.1} & 6.4 & 70.4 & 15.9 & 72.9 & 14.7 \\
& WPMIA & \textbf{90.2} & \textbf{57.8} & \textbf{86.5} & \underline{35.9} & \textbf{74.3} & \textbf{23.9} & \textbf{86.9} & \textbf{45.4} & \textbf{84.5} & \textbf{40.8} \\
\midrule
\multirow{14}{*}{128} & \multicolumn{11}{c}{\cellcolor{gray!30}\texttt{Gray-Box}} \\
& Loss & 62.8 & 13.5 & 65.2 & 15.4 & 68.9 & 21.2 & 71.5 & 24.0 & 67.1 & 18.5\\
& Reference & 64.3 & 12.5 & 63.8 & 15.4 & 63.8 & 13.5 & 69.7 & 23.1 & 65.4 & 16.1\\
& Lowercase & 61.2 & 6.7 & 64.8 & 3.8 & 67.6 & 20.2 & 68.0 & 8.7 & 65.4 & 9.9\\
& Zlib & 63.6 & 16.3 & 66.9 & 18.3 & 70.2 & 20.2 & 72.3 & 25.0 & 68.3 & 20.0\\
& Neighbor & 64.0 & 16.3 & 69.3 & 13.5 & 70.3 & 22.1 & 73.6 & 22.1 & 69.3 & 18.5\\
& Min-K\% & 67.6 & 15.4 & 69.5 & \underline{20.2} & 73.3 & 19.2 & 75.9 & 24.0 & 71.6 & 19.7 \\
& Min-K\%++ & \textbf{71.4} & \textbf{25.0} & \underline{72.2} & 19.2 & \underline{86.4} & \underline{38.5} & \underline{76.5} & \underline{28.9} & \underline{76.6} & \underline{27.9}\\
& ReCaLL & \underline{71.0} & \underline{17.3} & \textbf{87.4} & \textbf{41.3} & \textbf{91.3} & \textbf{48.1} & \textbf{88.0} & \textbf{39.4} & \textbf{84.4} & \textbf{36.5}\\
\cmidrule{2-12}
& \multicolumn{11}{c}{\cellcolor{gray!30}\texttt{Black-Box}} \\
& PETAL & 64.8 & 12.5 & 65.2 & \underline{19.2} & 64.3 & \underline{13.5} & 71.5 & 34.6 & 66.5 & 20.0\\
& SaMIA & 67.2 & 11.5 & 62.4 & 7.7 & \underline{65.2} & 7.7 & 67.5 & 16.3 & 65.6 & 10.8\\
& SimMIA$^*$ & \underline{85.5} & \underline{14.4} & \underline{82.8} & 16.3 & 61.7 & 9.6 & \underline{86.2} & \underline{40.4} & \underline{79.0} & \underline{20.1} \\
& SimMIA & 78.9 & 13.5 & 73.4 & 9.6 &  60.0 & 3.8 & 71.3 & 20.2 & 70.9 & 11.8 \\
& WPMIA & \textbf{87.2} & \textbf{49.0} & \textbf{84.7} & \textbf{39.4} & \textbf{67.4} & \textbf{15.4} & \textbf{87.5} & \textbf{48.1} & \textbf{81.7} & \textbf{38.0} \\
\bottomrule
\end{tabular}
}
\caption{Results on the WikiMIA benchmark. \textbf{Bold} and \underline{underlined} values indicate the best and second-best results in each column, respectively. $^*$ denotes scoring with \eqnref{eq:simmia_hard} instead of \eqnref{eq:simmia_soft}.}
\label{tab:wikimia}
\vspace{-10pt}
\end{table*}

%% file: tabs/mimir.tex
\begin{table*}[t!]
\centering
\resizebox{\linewidth}{!}{
\setlength{\tabcolsep}{5pt}
\begin{tabular}[t]{lcccccccccccccccc}
\toprule
\makecell[c]{\multirow{2}{*}{\textbf{Method}}} & \multicolumn{4}{c}{\textbf{Wikipedia}} & \multicolumn{4}{c}{\textbf{Github}} & \multicolumn{4}{c}{\textbf{Pile CC}} & \multicolumn{4}{c}{\textbf{PubMed Central}} \\
\cmidrule(lr){2-5}\cmidrule(lr){6-9}\cmidrule(lr){10-13}\cmidrule(lr){14-17}
& 160M & 1.4B & 2.8B & 6.9B & 160M & 1.4B & 2.8B & 6.9B & 160M & 1.4B & 2.8B & 6.9B & 160M & 1.4B & 2.8B & 6.9B \\
\midrule
\rowcolor{gray!30}
\multicolumn{17}{c}{\texttt{Gray-Box}}\\
Loss & \textbf{62.3} & \textbf{64.9} & \underline{66.2} & \underline{66.4} & 83.6 & 86.4 & 88.4 & 87.9 & \textbf{53.7}& 55.2 & \underline{54.9} & 57.0 & \underline{79.1} & \textbf{78.0} & \textbf{77.8} & \textbf{77.6}\\
Reference & 51.7 & 62.9 & \textbf{67.6} & \textbf{67.8} & 67.7 & 73.3 & 73.7 & 65.7 & 52.0 & \textbf{58.6} & \textbf{58.2} & \textbf{63.7} & 68.0 & 66.8 & 63.0 & 60.5\\
Lowercase & 58.6 & 62.1 & 63.5 & 64.1 & 76.5 & 81.2 & 83.5 & 83.3 & \underline{53.6} & 54.3 & 54.2 & 55.7 & 75.1 & 75.2 & 75.4 & 75.2\\
Zlib & 56.5 & 60.9 & 63.0 & 63.0 & \textbf{87.9} & \textbf{89.8} & \textbf{91.2} & \textbf{90.9} & 51.9 & 53.9 & 53.7 & 55.6 & 77.8 & 76.9 & \underline{76.7} & \underline{76.7}\\
Neighbor & 57.9 & 60.9 & 61.8 & 61.7 & \underline{86.6} & \underline{88.8} & \underline{90.5} & \underline{89.4} & 52.7 & 55.2 & 54.8 & 56.8 & 76.8 & 74.4 & 74.0 & 74.4\\
Min-K\% & 59.9 & \underline{63.8} & 65.6 & 65.8 & 82.9 & 86.5 & 88.4 & 88.3 & 52.9 & \underline{55.3} & 54.7 & 56.9 & 77.3 & \underline{77.4} & \textbf{77.8} & \textbf{77.6}\\
Min-K\%++ & 55.2 & 62.1 & 64.2 & 63.8 & 72.5 & 81.6 & 85.0 & 84.9 & 51.6 & 54.9 & 54.1 & 56.3 & 62.3 & 63.0 & 66.0 & 67.5\\
ReCaLL & 55.9 & 63.4 & 65.1 & 65.2 & 83.0 & 86.0 & 87.9 & 87.4 & 51.3 & 54.7 & 53.7 & \underline{57.4} & \textbf{79.5} & 75.2 & 76.0 & 73.7 \\
\midrule
\rowcolor{gray!30}
\multicolumn{17}{c}{\texttt{Black-Box}}\\
PETAL & 60.1 & 62.4 & 63.7 & 63.6 & 68.2 & 70.7 & 72.1 & 71.3 & \textbf{52.8} & \underline{53.9} & 54.0 & 55.4 & \underline{72.1} & \underline{72.7} & 73.3 & \underline{73.4}\\
SaMIA & 49.1 & 51.3 & 51.9 & 52.1 & 53.6 & 60.3 & 61.1 & 62.7 & 50.1 & 48.9 & 48.3 & 49.9 & 53.1 & 51.2 & 54.8 & 52.6\\
SimMIA$^*$ & 57.3 & 62.6 & 61.1 & 62.8 & 79.8 & 78.3 & 79.6 & 76.5 & 47.8 & 51.3 & 53.6 & 53.3 & 48.2 & 52.4 & 63.7 & 58.2\\
SimMIA & \underline{61.8} & \underline{64.3} & \underline{65.1} & \underline{65.0} & \underline{83.5} & \underline{85.4} & \underline{88.9} & \underline{85.1} & \textbf{52.8} & \textbf{55.1} & \underline{55.3} & \textbf{57.6} & 71.3 & 71.7 & \underline{75.1} & 65.1 \\
WPMIA & \textbf{62.4} & \textbf{64.8} & \textbf{65.7} & \textbf{65.5} & \textbf{87.4} & \textbf{88.3} & \textbf{90.6} & \textbf{88.2} & \underline{52.6} & \textbf{55.1} & \textbf{55.5} & \underline{56.9} & \textbf{74.5} & \textbf{73.9} & \textbf{77.9} & \textbf{74.2} \\
\toprule
\makecell[c]{\multirow{2}{*}{\textbf{Method}}} & \multicolumn{4}{c}{\textbf{ArXiv}} & \multicolumn{4}{c}{\textbf{DM Mathematics}} & \multicolumn{4}{c}{\textbf{HackerNews}} & \multicolumn{4}{c}{\textbf{Average}} \\
\cmidrule(lr){2-5}\cmidrule(lr){6-9}\cmidrule(lr){10-13}\cmidrule(lr){14-17}
& 160M & 1.4B & 2.8B & 6.9B & 160M & 1.4B & 2.8B & 6.9B & 160M & 1.4B & 2.8B & 6.9B & 160M & 1.4B & 2.8B & 6.9B \\
\midrule
\rowcolor{gray!30}
\multicolumn{17}{c}{\texttt{Gray-Box}}\\
Loss & 74.4 & \textbf{77.6} & \textbf{78.0} & \textbf{78.4} & \underline{94.6} & \underline{93.3} & 92.8 & \underline{92.9} & \underline{57.9} & \underline{59.3} & \textbf{60.5} & \textbf{60.7} & \textbf{72.2} & \textbf{73.5} & \textbf{74.1} & \textbf{74.4}\\
Reference & 57.6 & 71.6 & 71.5 & 71.9 & 61.2 & 48.3 & 44.9 & 44.9 & 51.5 & 53.7 & 57.1 & 57.4 & 58.5 & 62.2 & 62.3 & 61.7\\
Lowercase & 69.2 & 74.0 & 75.5 & 75.8 & 80.7 & 75.9 & 78.2 & 72.6 & 57.7 & 59.0 & 59.9 & \underline{60.2} & 67.3 & 68.8 & 70.0 & 69.6\\
Zlib & \underline{74.5} & \underline{77.2} & 77.7 & \underline{78.0} & 80.9 & 81.0 & 81.2 & 81.0 & 57.5 & 58.4 & 59.3 & 59.4 & 69.6 & 71.2 & 71.8 & 72.1\\
Neighbor & 70.8 & 74.9 & 75.5 & 76.1 & 79.7 & 71.4 & 73.2 & 71.5 & 56.7 & 57.5 & 58.4 & 58.2 & 68.7 & 69.0 & 69.7 & 69.7\\
Min-K\% & 68.6 & 74.2 & 65.3 & 75.6 & 93.1 & \underline{93.3} & \underline{92.9} & \textbf{93.0} & 55.0 & 56.9 & 57.9 & 58.6 &70.0 & 72.5 & 71.8 & \underline{73.7}\\
Min-K\%++ & 52.4 & 64.1 & 65.5 & 65.0 & 76.9 & 76.0 & 70.2 & 73.5 & 53.7 & 56.0 & 57.4 & 58.9 & 60.7 & 65.4 & 66.1 & 67.1\\
ReCaLL & \textbf{76.5} & 76.5 & \underline{77.9} & 76.5 & \textbf{96.7} & \textbf{94.4} & \textbf{94.2} & 91.8 & \textbf{59.1} & \textbf{59.5} & \underline{60.1} & \textbf{60.7} & \underline{71.7} & \underline{72.8} & \underline{73.6} & 73.2\\
\midrule
\rowcolor{gray!30}
\multicolumn{17}{c}{\texttt{Black-Box}}\\
PETAL & 68.4 & 70.9 & \underline{71.8} & 72.4 & 86.5 & 86.9 & 86.5 & 86.8 & \textbf{57.1} & \textbf{58.0} & \textbf{58.5} & \textbf{58.4} & 66.5 & 67.9 & 68.6 & 68.8\\
SaMIA & 43.6 & 44.6 & 45.4 & 45.9 & 82.3 & 85.3 & 79.9 & 73.4 & 49.3 & 49.6 & 49.5 & 50.1 & 54.4 & 55.9 & 55.8 & 55.2\\
SimMIA$^*$ & 57.1 & 57.9 & 54.3 & 61.5 & 64.4 & 65.0 & 67.5 & 62.4 & 51.9 & 54.8 & 54.4 & 54.8 & 58.1 & 60.3 & 62.0 & 61.4 \\
SimMIA & \underline{71.7} & \underline{72.6} & 66.8 & \underline{72.5} & \underline{86.7} & \underline{89.6} & \underline{89.3} & \underline{90.3} & 53.1 & 53.3 & 52.8 & 53.0 & \underline{68.7} & \underline{70.3} & \underline{70.5} & \underline{69.8} \\
WPMIA & \textbf{74.4} & \textbf{75.6} & \textbf{73.6} & \textbf{75.3} & \textbf{91.6} & \textbf{94.9} & \textbf{94.3} & \textbf{95.3} & \underline{54.7} & \underline{55.6} & \underline{56.1} & \underline{56.0} & \textbf{71.1} & \textbf{72.6} & \textbf{73.4} & \textbf{73.1} \\
\bottomrule
\end{tabular}
}
\caption{AUC results on the MIMIR benchmark. \textbf{Bold} and \underline{underlined} values indicate the best and second-best results in each column, respectively. $^*$ denotes scoring with \eqnref{eq:simmia_hard} instead of \eqnref{eq:simmia_soft}.}
\label{tab:mimir}
\vspace{-10pt}
\end{table*}

%% file: tabs/wikimia-25.tex
\begin{table*}[t!]
\centering
\resizebox{\linewidth}{!}{
\setlength{\tabcolsep}{5pt}
\begin{tabular}{lcccccccccc}
\toprule
\makecell[c]{\multirow{2}{*}{\textbf{Method}}} &
\multicolumn{2}{c}{\textbf{Pythia-6.9B}} &
\multicolumn{2}{c}{\textbf{Qwen3-8B-Base}} &
\multicolumn{2}{c}{\textbf{Claude-4.5-Haiku}} &
\multicolumn{2}{c}{\textbf{Gemini-2.5-Flash}} &
\multicolumn{2}{c}{\textbf{GPT-5-Chat}} \\
\cmidrule(lr){2-3}\cmidrule(lr){4-5}\cmidrule(lr){6-7}\cmidrule(lr){8-9}\cmidrule(lr){10-11}
& AUC & TPR@5\%FPR & AUC & TPR@5\%FPR & AUC & TPR@5\%FPR & AUC & TPR@5\%FPR & AUC & TPR@5\%FPR \\
\midrule
\multicolumn{11}{>{\columncolor{gray!30}}c}{\texttt{Gray-Box}} \\
Loss & 73.0 & 36.3 & 59.1 & 11.5 & - & - & - & - & - & -\\
Reference & 65.2 & 8.9 & 53.9 & 3.5 & - & - & - & - & - & -\\
Lowercase & 70.7 & 23.0 & 59.4 & 19.5 & - & - & - & - & - & -\\
Zlib & 72.2 & 31.9 & 57.5 & 8.9 & - & - & - & - & - & -\\
Min-K\% & \underline{77.3} & 38.9 & 56.8 & 14.2 & - & - & - & - & - & -\\
Min-K\%++ & 72.7 & \underline{41.6} & \underline{62.7} & \underline{20.4} & - & - & - & - & - & -\\
ReCaLL & \textbf{96.1} & \textbf{78.8} & \textbf{91.0} & \textbf{43.4} & - & - & - & - & - & -\\
\midrule
\multicolumn{11}{>{\columncolor{gray!30}}c}{\texttt{Black-Box}} \\
PETAL & 72.2 & 30.1 & 62.8 & \underline{18.6} & - & - & 67.1 & 23.9 & 78.0 & 35.4\\
SaMIA  & 57.8 & 8.8 & 50.3 & 2.7 & 62.6 & \underline{8.0} & 57.0 & 15.0 & 65.7 & 13.3 \\
SimMIA$^*$ & 88.1 & 37.2 & 63.2 & 2.7 & \underline{65.7} & 6.2 & \underline{75.7} & \underline{30.2} & \underline{90.6} & \textbf{73.3}\\
SimMIA & \underline{88.8} & \underline{47.8} & \underline{63.0} & 6.2 & 60.5 & 6.2 & 74.1 & 28.5 & 83.3 & 37.2\\
WPMIA & \textbf{89.0} & \textbf{56.6} & \textbf{65.9} & \textbf{23.0} & \textbf{66.3} & \textbf{22.1} & \textbf{78.2} & \textbf{36.5} & \textbf{91.0} &  \underline{67.3}\\
\bottomrule
\end{tabular}
}
\caption{Results on the WikiMIA-25 benchmark. \textbf{Bold} and \underline{underlined} values indicate the best and second-best results in each column, respectively. $^*$ denotes scoring with \eqnref{eq:simmia_hard} instead of \eqnref{eq:simmia_soft}.}
\label{tab:wikimia25}
\vspace{-10pt}
\end{table*}